\documentclass{elsart}
\usepackage{amssymb}
\usepackage{graphicx}

\begin{document}

\newcommand\cc{{\rm CC}}
\newcommand\nutau{{\nu_\tau}}
\newcommand\anutau{\bar{\nu}_\tau}
\newcommand\numu{{\nu_\mu}}
\newcommand\anumu{\bar{\nu}_\mu}
\newcommand\nue{{\nu_e}}
\newcommand\anue{\bar{\nu}_e}

\newcommand\enunu{e^- \anue \nutau}
\newcommand\mununu{\mu^- \anumu \nutau}
\newcommand\taue{\tau^- \to \enunu}
\newcommand\taumu{\tau^- \to \mununu}
\newcommand\tauonepr{\tau^- \to h^- (n\pi^0) \nutau}
\newcommand\taupi{\tau^- \to \pi^- \nutau}
\newcommand\taurho{\tau^- \to \rho^- \nutau}
\newcommand\threepi{\pi^-\pi^+\pi^-}
\newcommand\tauthreepi{\tau^- \to \threepi \nutau}
\newcommand\tauthreepr{\tau^- \to \threepi (n\pi^0) \nutau}
\newcommand\tauplusthreepi{\tau^+ \to \pi^+\pi^-\pi^+ \anutau}

\newcommand\numunutau{\numu \to \nutau}
\newcommand{\ra}{\rightarrow}
\newcommand{\nmnt}{$\numu \ra \nutau \; $}
\newcommand{\nmns}{$\numu \ra \nu_s \; $}
\newcommand{\nmne}{$\numu \ra \nue \; $}
\newcommand{\emurat}{\nue/\numu}
\newcommand{\Dms}{\Delta m^2}

\newcommand\epsmu{\epsilon_\mu}
\newcommand\epstau{\epsilon_\tau}
\newcommand\Eetot{E_{\rm e}^{\rm tot}}
\newcommand\evis{E_{\rm vis}}
\newcommand\qlep{Q_{\rm Lep}}
\newcommand\pt{p_T}
\newcommand\xvis{\tau_{\rm V}}
\newcommand\vpxvis{\vec p^{\,\xvis}}
\newcommand\jet{H}                    

\newcommand{\pip}{\pi^+}
\newcommand{\pim}{\pi^-}
\newcommand{\pis}{\pi^\pm}
\newcommand{\kp}{K^+}
\newcommand{\km}{K^-}
\newcommand{\ks}{K^\pm}
\newcommand{\klong}{K^0_L}
\newcommand{\kshort}{K^0_S}
\newcommand{\mup}{\mu^+}
\newcommand{\mum}{\mu^-}
\newcommand{\mus}{\mu^\pm}

\newcommand\epsdir{.}

\newcommand{\fix}[1]{{\bf <<< #1 !!! }}

\begin{frontmatter}

\title{Prediction of neutrino fluxes in the NOMAD experiment}
{\large NOMAD Collaboration}

\author[Paris]             {P.~Astier},
\author[CERN]              {D.~Autiero},
\author[Saclay]            {A.~Baldisseri},
\author[Padova]            {M.~Baldo-Ceolin},
\author[Paris]             {M.~Banner},
\author[LAPP]              {G.~Bassompierre},
\author[Lausanne]          {K.~Benslama},
\author[Saclay]            {N.~Besson},
\author[CERN,Lausanne]     {I.~Bird},
\author[Johns Hopkins]     {B.~Blumenfeld},
% ...
\corauth[corr]                {Corresponding author.}
\ead                          {Slava.Valouev@cern.ch}
\author[Paris]             {F.~Vannucci},
\author[Sydney]            {K.E.~Varvell},
\author[Urbino]            {M.~Veltri},
\author[Pavia]             {V.~Vercesi},
\author[CERN]              {G.~Vidal-Sitjes},
\author[Lausanne]          {J.-M.~Vieira},
\author[UCLA]              {T.~Vinogradova},
\author[Harvard,CERN]      {F.V.~Weber},
\author[Dortmund]          {T.~Weisse},
\author[CERN]              {F.F.~Wilson},
\author[Melbourne]         {L.J.~Winton},
\author[Sydney]            {B.D.~Yabsley},
\author[Saclay]            {H.~Zaccone},
\author[Dortmund]          {K.~Zuber}

\address[LAPP]           {LAPP, Annecy, France}
\address[Johns Hopkins]  {Johns Hopkins Univ., Baltimore, MD, USA}
\address[Harvard]        {Harvard Univ., Cambridge, MA, USA}
% ...
\address[Dortmund]       {Dortmund Univ., Dortmund, Germany}
% ...
\address[CERN]           {CERN, Geneva, Switzerland}
\address[Lausanne]       {University of Lausanne, Lausanne, Switzerland}
\address[UCLA]           {UCLA, Los Angeles, CA, USA}
\address[Melbourne]      {University of Melbourne, Melbourne, Australia}
% ...
\address[Padova]         {Univ. of Padova and INFN, Padova, Italy}
\address[Paris]          {LPNHE, Univ. of Paris VI and VII, Paris, France}
\address[Pavia]          {Univ. of Pavia and INFN, Pavia, Italy}
% ...
\address[Saclay]         {DAPNIA, CEA Saclay, France}
% ...
\address[Sydney]         {Univ. of Sydney, Sydney, Australia}
\address[Urbino]         {Univ. of Urbino, Urbino, and INFN Florence, Italy}
% ...

\begin{abstract}
The method developed for the calculation of the flux and composition of the
West Area Neutrino Beam used by NOMAD in its search for neutrino
oscillations is described.  The calculation is based on particle production
rates computed using a recent version of FLUKA and modified to take into
account the cross sections measured by the SPY and NA20 experiments.  These
particles are propagated through the beam line taking into account the
material and magnetic fields they traverse.  The neutrinos produced through
their decays are tracked to the NOMAD detector.  The fluxes of the four
neutrino flavours at NOMAD are predicted with an 
uncertainty of about 8\% for $\numu$ and $\nue$, 10\% for $\anumu$,
and 12\% for $\anue$.  The energy-dependent uncertainty achieved on
the $\emurat$ prediction needed for a \nmne oscillation search ranges
from 4\% to 7\%, whereas the overall normalization uncertainty on this
ratio is 4.2\%.
\end{abstract}


\end{frontmatter}

\vspace*{-0.2cm}
\section{Introduction}
\label{sec:intro}
 The NOMAD experiment~\cite{NOMADNIM,NUTAU_LATEST} is searching for \nmnt
 and \nmne oscilla\-tions in a predominantly $\numu$ beam at CERN.
 Oscillations between $\numu$ and $\nue$ would be evidenced
 by a distortion of the energy distribution of the intrinsic $\nue$
 component of the beam. The \nmnt search requires the understanding
 of the major component, $\numu$, of the beam in order to interpret
 any potential oscillation signal, and of the minor components of the beam
 in order to calculate reliably various backgrounds. Thus it is imperative
 to understand the composition of the beam.  In addition, to search for
 \nmne oscillations, a prediction of a $\emurat$ ratio and the understanding
 of its systematic uncertainty is crucial.

 The beam was produced through the decay of mesons originating
 in the interaction of protons with a beryllium target. This paper
 describes one of the two methods used to predict the neutrino flux at NOMAD
 and the performance of the beam. One of the most critical ingredients
 in the simulation program used to describe the beam is the set of particle
 production cross sections assumed in the initial p-Be interaction.
 Given the paucity of data on these cross sections, a dedicated
 experiment~\cite{SPY}, referred to as SPY, in which some members
 of the NOMAD collaboration participated, was performed at the CERN
 450~GeV proton synchrotron (SPS).  It measured charged particle yields
 in the relevant energy and angular regions.

 Two distinct methods were then used to predict secondary particle production
 as input to the simulation. The first used particle yields
 from a recent version of FLUKA~\cite{FLUKA} suitably
 corrected to take into account the SPY results, and is described in this
 paper.  The second, referred
 to as the Empirical Parameterization (EP) method, was used to predict
 the $\nue$ flux.  It used the NOMAD $\numu$, $\anumu$ and $\anue$ flux
 data to estimate the $\mup$, $\kp$ and $\klong$ production rates at the target
 and thus predict the $\nue$ rate. It also used the SPY data to constrain
 the low energy $\kp$ rates as well as a measurement by Skubic et
 al.~\cite{Skubic} to constrain the $K^0_L$ contribution.  The EP method
 will be described in a forthcoming publication.

 This publication is structured as follows. Section~\ref{sec:beamhard}
 describes the neutrino beam hardware, Section~\ref{sec:control} its
 monitoring and alignment, Section~\ref{sec:yields} the particle production
 measurements used, Section~\ref{sec:mc} the simulation,
 Section~\ref{sec:fluka_comp} the beam composition, and
 Section~\ref{sec:fluka_syst} the systematic uncertainties.
 Section~\ref{sec:NOMAD} briefly describes the NOMAD apparatus and running
 conditions,
 Section~\ref{sec:compData} compares the results of our simulations with
 the data collected in NOMAD, Section~\ref{sec:ratio} presents the final
 $\emurat$ predictions, and Section~\ref{sec:conclusions} draws some
 conclusions.
 
\section{Beam description}
\label{sec:beamhard}
 The neutrino beam is produced by extracting part of the 450~GeV proton
 beam circulating in the SPS and letting it interact with a beryllium target.
 Positively charged particles (mainly $\pip$ and $\kp$ mesons) produced
 around zero degrees with
 respect to the primary proton beam are focused into a near parallel beam
 by a system of magnetic lenses and subsequently decay producing neutrinos.
 A large iron and earth shield placed at the end of the decay volume filters
 out particles other than neutrinos and is followed by the detectors,
 CHORUS~\cite{CHORUS} and NOMAD.

\begin{figure}[tbhp]
\vspace*{0.8cm}
\begin{center}
  \includegraphics*[width=13.9cm]{\epsdir/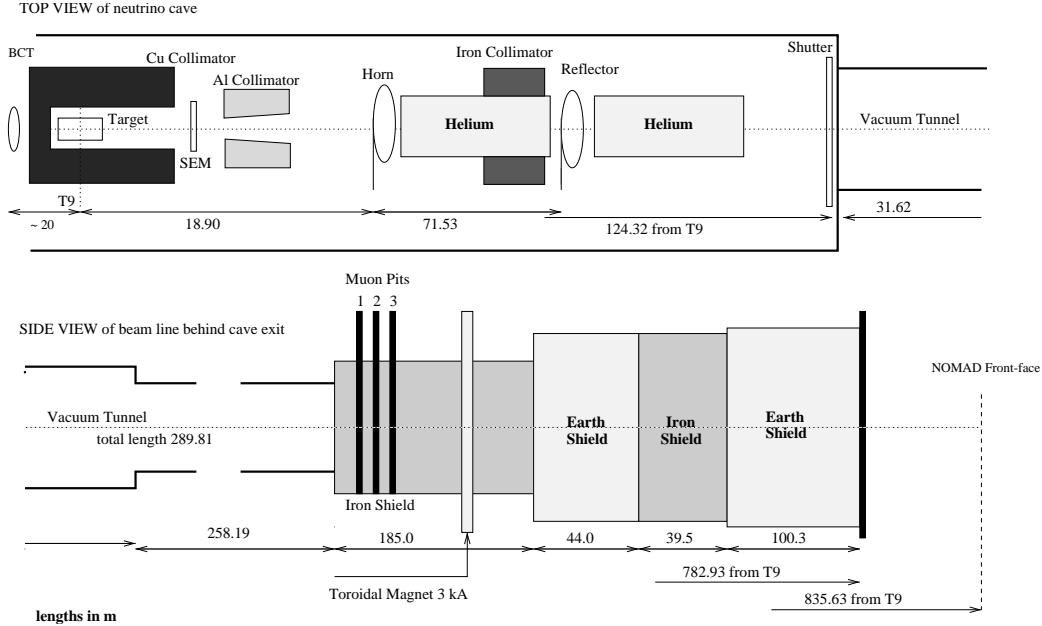}
\end{center}
\caption{Schematic layout of the WANF beam line (not drawn to scale).}
\label{fig:beam_wanf}
\end{figure}

 The general layout of the beam line, referred to as the West Area Neutrino
 Facility (WANF), is illustrated in Fig.~\ref{fig:beam_wanf}. The beam line
 operated for more than 20 years and was re-optimized~\cite{WANF}
 in 1992 and 1993 for the NOMAD and CHORUS experiments.

\subsection{The proton beam}
\label{sec:protons}
 During the lifetime of the CHORUS and NOMAD experiments the SPS
 accelerated up to $4.3 \times 10^{13}$ protons per 14.4~s cycle.
 The protons used to produce neutrinos were ejected from the SPS through
 resonant extraction in two spills, one towards the end of the
 accelerating ramp at 445~GeV and the second, 2.7~s later, at the end
 of the 450~GeV flat top. Each of these two spills had a full width
 at half maximum of 3~ms and contained about
 $1.8 \times 10^{13}$ protons. The length of the spill was dictated
 by the requirement to keep the detector live time above 90\% while
 remaining compatible with the maximum possible duration of the current
 pulses in the focusing magnets (Section~\ref{sec:lenses}).

\subsection{The target}
\label{sec:target}
 The target consisted of 11 beryllium rods separated by 90~mm, each 3~mm
 in diameter and 100~mm long. The rods were supported by beryllium disks
 and enclosed in an aluminium target box. Gaseous helium under pressure
 was directed at each rod for cooling purposes. The entrance and exit
 windows of the box consisted of 60~mm diameter titanium foils each 0.1~mm
 thick. Each of the two extremities of the box could be displaced laterally
 by $\pm 12$~mm for alignment purposes. The target box was surrounded
 by iron and marble shielding slabs and along the beam direction
 by collimators.

 The target amounted to 2.7 nuclear interaction lengths resulting in
 only 6.7\% of the protons not undergoing inelastic interactions in it.
 Since the SPS is installed in an underground tunnel and the detectors
 were located on the surface, the primary proton beam pointed upwards
 at an angle of 42~mrad and the target box was located in an underground
 area at a depth of 35~m.

\subsection{The collimators}
\label{sec:collimators}
 The target was immediately followed by a copper collimator 1.20~m long
 with an 85~mm cylindrical bore, followed by an
 aluminium collimator, 2.75~m long starting 3.55~m from the centre of the
 target. The aluminium collimator defined an average angular acceptance
 of 10~mrad for secondaries produced at the target. Both collimators
 were water cooled.

\subsection{The magnetic horn and reflector}
\label{sec:lenses}
\begin{figure}[b]
\begin{center}
  \includegraphics*[scale=0.77]{\epsdir/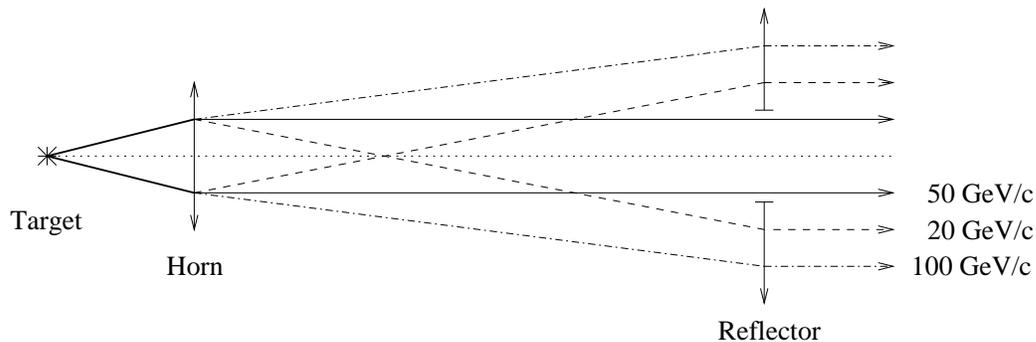}
\end{center}
\caption{Principle of the focusing.  The lines are representative
  trajectories of particles of three different momenta.}
\label{fig:focusing}
\end{figure}

 Two toroidal magnetic lenses, referred to as the horn and the reflector,
 focused charged particles of a given sign (positive for a predominantly
 $\numu$ beam) produced
 at the target into a near parallel beam while defocusing the particles
 of the opposite charge. The principle of the focusing is illustrated
 in Fig.~\ref{fig:focusing}. The reflector provided additional focusing
 for high momentum particles and compensated for overfocusing of low
 momentum particles by the horn. The magnetic field was provided by
 current sheets flowing in the inner and outer conductors of the lenses.
 The field was measured to be azimuthally symmetric to better than 1.5\%.
 Its value at a
 radial position $r$ from the beam axis and for a current $I$ is given by
\begin{equation}
 B = \mu_\circ I/2\pi r \label{eq:bfield}.
\end{equation}
 The current (100~kA for the horn and 120~kA for the reflector) was provided
 by the discharge of capacitor banks and lasted 6.8~ms. The thickness of
 the inner conductors was minimized to reduce secondary interactions
 while maintaining adequate strength to withstand the magnetic forces.
 Both elements were made of aluminium alloys of various tensile strengths.

 The inner conductor of the 6.56~m long horn was approximately parabolical
 in shape. At its upstream end it had a diameter of 140~mm and a thickness
 of 1.7~mm while at the downstream end its diameter decreased to 8~mm
 and its thickness was therefore increased to 6.8~mm in order to
 withstand the mechanical stress caused by strong magnetic fields.
 The outer conductor consisted of a 420~mm diameter cylinder of 12~mm
 thickness. The conductors were built in sections joined together by flanges.
 The centring of the inner conductor was achieved with steel cables joining
 the inner and outer conductors through insulating spacers.

 The 6.54~m long reflector had an inner conductor diameter of 416~mm
 decreasing to 196~mm, and an outer conductor of 776~mm diameter. The inner
 conductors of both elements were cooled by spraying water onto them from
 nozzles located at the top of the outer conductor.

 The positions of the horn and of the reflector, 18.9~m and 90.4~m
 from the centre of the target, were chosen to provide a high energy
 neutrino spectrum best suited to the detection of $\nutau$ charged
 current ($\cc$) interactions. An iron collimator placed between the horn
 and the reflector absorbed particles of the wrong charge swept away by
 the horn before their decay thus reducing the contamination of
 antineutrinos in the neutrino beam.

 The polarity of these magnetic elements could be changed within minutes in
 order to produce an antineutrino beam.

\subsection{The helium bags}
\label{sec:hebags}
 Two helium bags were installed to reduce multiple scattering and secondary
 interactions along the beam: one, 63~m long, between the horn and the
 reflector
 and the other, 18~m long, between the reflector and the decay tunnel.
 Each bag was closed by 0.3~mm thick titanium windows. Replacing the air
 by these helium bags resulted in a 7\% increase in the neutrino flux.

\subsection{Ionization chambers}
\label{sec:chambers}
 Two cylindrical ionization chambers, 840~mm and 60~mm in diameter, were
 added to the beam line before the 1996 and 1998 runs respectively.
 Their purpose was to measure the flux and
 profiles of secondary particles and of protons that did not interact
 in the target.
 The chambers were placed between the second helium bag and the entrance
 to the decay tunnel.

\subsection{The decay tunnel}
\label{sec:dtun}
 A 289.9~m long tunnel was provided to allow a significant fraction of
 the $\pip$ and $\kp$ to decay. This decay tunnel was evacuated to a pressure
 of 10~Torr. It consisted of a 31.6~m long section of 2.2~m diameter
 followed by a 258.3~m long section of 1.2~m diameter. The entrance window
 to the tunnel was made of 2~mm thick titanium.

 The decay tunnel contained hardware that had only been used in earlier
 experiments to position an absorber in the tunnel in order to enhance
 the fraction of prompt neutrinos in the beam by reducing the number
 of neutrinos originating from long lived particles that decayed in the
 tunnel. This hardware was located near the outer radius of the decay
 tunnel, at its downstream end, and was not used during the CHORUS and
 NOMAD runs.

\subsection{The hadron and muon filter}
\label{sec:filters}
 The very large flux of hadrons and muons emerging at the end of the
 decay tunnel had to be absorbed before the detectors. This was accomplished
 by a shield consisting of two iron filters, one 185~m long and the other
 39.5~m long, separated by 44.0~m of earth and followed by an additional
 100.3~m of earth. A 10~m long magnetized iron toroid with the field of
 1~T was embedded in the
 front iron filter. It had a 2~m inner diameter and a 6~m outer diameter
 and its purpose was to deflect muons away from the beam direction.

\section{Monitoring and alignment}
\label{sec:control}
 The monitoring of the neutrino beam involved the measure of the proton
 intensity on target, the centring of the beam relative to the target,
 the width of the beam pulse and its timing relative to the horn current
 pulse.  All the elements of the neutrino beam line were aligned with
 respect to the incident proton beam before the start of data taking
 in the CHORUS and NOMAD experiments.

\subsection{Monitoring of the incident proton beam}
\label{sec:monitoring}
 The proton flux was measured with two beam current transformers, one located
 immediately after the extraction from the SPS and the other before the
 target. A secondary emission monitor (SEM) placed just upstream of the
 target also yielded a measure of the proton intensity.

\begin{figure}
\begin{center}
  \includegraphics*[scale=0.65]{\epsdir/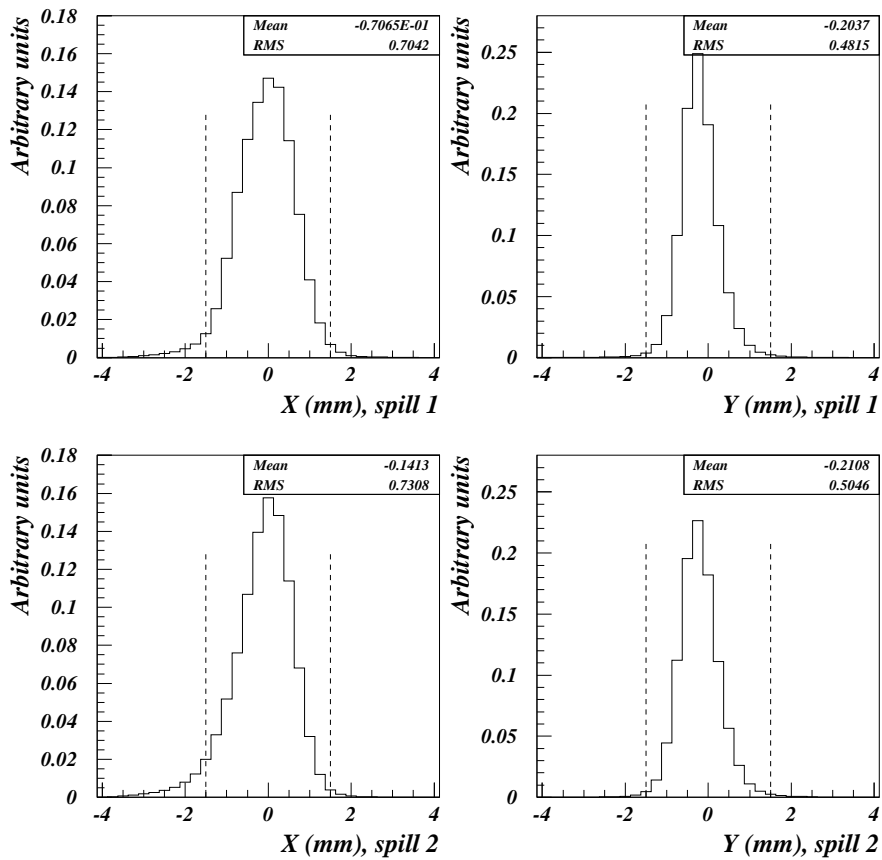}
\end{center}
\vspace{-0.5cm}
\caption{Measured profiles of the incident proton beam, for both spills,
  averaged over all years of data taking.  The dashed vertical lines indicate
  the edges of the target.}
\label{fig:prot_prof}
\end{figure}

 The alignment of the beam with respect to the target was checked by
 measuring the ratio of pulse heights in two SEM's, one downstream of
 the target and one upstream~\cite{SEM}.  This ratio, referred to as
 the multiplicity,
 decreased if the beam was not centred because of fewer secondary
 particles being produced and reaching the downstream SEM. The centring
 and width of the beam on target was also checked periodically with
 a horizontal and a vertical beam scanner each consisting of a wire
 moved in steps across the beam just in front of the target from $-4$~mm
 to $+4$~mm. The measured profiles (Fig.~\ref{fig:prot_prof}) were used
 as input to the simulation program. Their typical full width at half
 maximum was 1.7~mm in $X$ and 1.0~mm in $Y$. The beam was well
 contained within the 3 mm diameter target and only 5.2\% of the protons
 missed it.

 Further checks were provided by secondary emission monitors consisting
 of foils split in two halves either in the vertical or in the horizontal
 direction and placed just beyond the target. Equality of the secondary
 particle flux in the left, right halves and in the up, down halves
 ensured centring of the beam.

 A visual oscilloscope display provided a monitoring of the width of
 the beam spill and of its correct timing relative to the horn and
 reflector pulses. Furthermore, a narrowing of the beam pulse resulted
 in an increased experimental dead time and generated an alarm.

\subsection{Monitoring of the muon flux}
\label{sec:muflux}
 An overall measure of the stability of the neutrino flux intensity
 and direction as well
 as of the performance of the horn and reflector was provided by studying
 the accompanying muon flux. The muon flux was monitored by 3 planes
 (V1, V2 and V3) of solid state diodes (SSD) positioned within the
 first iron filter in pits located after 10.4~m, 30.8~m and 50.8~m
 of iron~\cite{SSD}.
 Planes V1, V2 and V3 consisted respectively of 19, 14 and 10 SSD's
 fixed in positions such as to sample the radial and azimuthal
 distributions of the muon flux. In each plane a movable calibration box
 containing 5 additional counters provided an inter-counter calibration.
 A reference box could be moved from pit to pit for inter-plane calibration.
 The charge deposited in each SSD was recorded for each spill, thus
 providing an on-line measure of the stability of the muon, and therefore
 of the neutrino, flux as well as of its direction.


\subsection{Alignment}
\label{sec:alignment}
 The goal of the alignment
 exercise~\cite{align} was to maximize the neutrino flux
 and centre it as well as possible using as control the various beam
 monitors available in the beam line and at the experiments.

 First the target was moved transversely to the beam in the vertical and
 horizontal directions, while keeping its length parallel to the beam.
 The optimal position was defined as the one
 yielding the largest multiplicity as defined in Section~\ref{sec:monitoring}.
 Movements of 2.8~mm horizontally and 0.3~mm vertically were necessary.
 The tilt of the target relative to the beam was also checked but was
 found to be correct.

 The position of the horn relative to the beam and target was
 then optimized by searching for the maximum value and best centring
 of the muon flux in the 3 pits. This was obtained for a relative
 displacement of 4.5~mm and 5.0~mm in the horizontal and vertical
 direction respectively, resulting in a 3\% increase of the muon flux
 and in a better centring of its spatial distribution by a few centimeters.

 This optimization was also checked by analyzing neutrino events in CHORUS
 and NOMAD. Shifts in the spatial distributions of events of about 10~cm
 were observed after optimization, resulting in the event distributions
 being in better agreement with those of events simulated with an ideal
 alignment. An 8\% increase in the event rate also resulted from this
 optimization~\cite{align}.

\section{Particle production measurements}
\label{sec:yields}

One of the most important ingredients in the calculation of the neutrino
flux and energy spectra presented in this paper were the results of two
measurements of the production rates of charged particles in p-Be
interactions.  These measurements were performed by the NA20 and the
NA56/SPY collaborations and covered complementary ranges of secondary
particle momenta, from 60 GeV/$c$ to 300 GeV/$c$ and from 7 GeV/$c$
to 135 GeV/$c$, respectively.

\subsection{The NA20 experiment}
\label{sec:Atherton}


The NA20 particle production experiment~\cite{Atherton} was performed
in the North Area of the CERN SPS using a 400 GeV/$c$ proton beam
incident on a variety of beryllium targets.  The relevant one for this
analysis was a plate 100 mm thick, 160 mm wide and 2 mm high.  Bending
magnets and collimators downstream of the target selected secondary
particles produced with the appropriate production angle, momentum and
charge.  Differential \v{C}erenkov counters (CEDARS) identified $\pi$,
$K$ and protons.  Production rates of charged secondary particles produced
with momentum of 60, 120, 200 and 300 GeV/$c$ were measured for two
values of their transverse momentum, 0.0 and 0.5 GeV/$c$.  At 120 GeV/$c$
the rates were also measured for a transverse momentum of 0.3 GeV/$c$.
The detailed results on the particle production rates, on the $K/\pi$
ratios and on the statistical and systematic uncertainties of these
measurements are given in Ref.~\cite{Atherton}, where they are
expressed as ``yields''.  The yield $Y$ is $\d^2 n/(\d\Omega \d p/p) \propto
p \d^2\sigma/(\d p \d \Omega)$, where $n$ is the number of observed
particles per incident proton.
At high energy, $Y$ is
proportional to $p^2$ times the Lorentz-invariant cross section
$E \d^3\sigma/\d^3p \approx (1/p) \d^2\sigma/(\d p \d \Omega)$.
In using the NA20 measurements, taken at 400 GeV/$c$
incident momentum, for particle production predictions at 450 GeV/$c$,
we assume the Feynman scaling hypothesis, i.e., that the
Lorentz-invariant cross section (expressed in terms of $p_T$ and the
Feynman variable $x_F$) is the same at these two beam momenta
$p_{\rm incident}$.  Then the prediction can be made at secondary
particle momenta having the same $x_F = p_{\rm L}/p_{\rm incident}$ at
450 GeV/$c$ as that at 400 GeV/$c$, if the yields are scaled up by
(450/400)$^2$.

\subsection{The NA56/SPY experiment}
\label{sec:SPY}
The NA56/SPY particle production experiment~\cite{SPY} was similar to
NA20 and was again performed in the North Area of the CERN SPS but
using a 450 GeV/$c$ proton beam.  It also used a variety of beryllium
targets but the relevant one for this analysis was identical to the
one described for NA20, namely a plate 100 mm thick, 160 mm wide and 2
mm high.  Bending magnets and collimators downstream of the target
selected particles of the appropriate production angle, momentum and
charge.  Time-of-flight counters, and threshold and differential
\v{C}erenkov counters identified $\pi$, $K$ and protons.  A calorimeter
separated electrons and muons from hadrons.  Data were collected at
0$^\circ$ production angle at 7, 10, 15, 20, 30, 40, 67.5, and 135 GeV/$c$
secondary particle momentum.  In addition several production angles,
ranging in transverse momentum from 0.0 to 0.6 GeV/$c$, were measured at
15 and 40 GeV/$c$.  The detailed results on the particle production
rates, on the $K/\pi$ ratios and on the statistical and systematic
uncertainties of these measurements are given in Ref.~\cite{SPY}.

\section{Beam simulation}
\label{sec:mc}
 A full Monte Carlo simulation of the WANF beam line has been performed.
 It used the measured profiles and the calculated divergences of the
 proton beam incident on the beryllium target as input and
 was implemented in two main steps.

 First, the yields of the secondary
 particles from p-Be interactions were calculated using FLUKA.
 FLUKA is a general purpose Monte Carlo package which contains,
 in particular, a detailed description of hadron-nucleon and
 hadron-nucleus interactions~\cite{FL_PHYS}. It is based on the
 Dual Parton Model~\cite{DPM} complemented by the simulation of
 nuclear reinteractions~\cite{INC}. It has been successfully tested
 over a variety of experimental data~\cite{FL_PHYS}.
 A recent version of FLUKA~\cite{FLUKA}, referred to as FLUKA 2000,
 was used (Section~\ref{sec:fluka}).  The FLUKA 2000 yields were corrected
 to take into account the results of the SPY and NA20 measurements
 (Section~\ref{sec:reweight}).  FLUKA 2000 was also used to
 transport the secondaries within the boundaries of the target box
 (Section~\ref{sec:reint_target}).

 At the second step, the secondaries were propagated
 up to the NOMAD detector (located 835~m from the target)
 using the NOMAD beam line simulation package, NUBEAM. It was based
 on GEANT~3.21~\cite{GEANT} and the 1992 version of FLUKA, GFLUKA,
 implemented within it~\cite{GFLUKA}. It therefore included the effects
 of energy losses, multiple scattering, reinteractions and decays.
 The GFLUKA meson yields were corrected by the corresponding ratios
 between FLUKA 2000 and GFLUKA.
 In this Section we describe the crucial points of the NUBEAM package
 -- the simulation of the magnetic field in the horn and reflector
 (Section~\ref{sec:B}), the simulation of the beam line hardware elements
 and the treatment of reinteractions (Section~\ref{sec:mat}), and the
 simulation of meson decays (Section~\ref{sec:decays}).

\subsection{FLUKA in NOMAD}
\label{sec:fluka}
One of the most critical elements in the prediction of neutrino
fluxes is the description of the yield of particles
in p-Be interactions.  In the approach used in this paper the yields
were obtained from a complete Monte Carlo simulation based on the FLUKA
generator modified to take into account available experimental data.

\begin{figure}
\begin{center}
  \includegraphics*[scale=0.48]{\epsdir/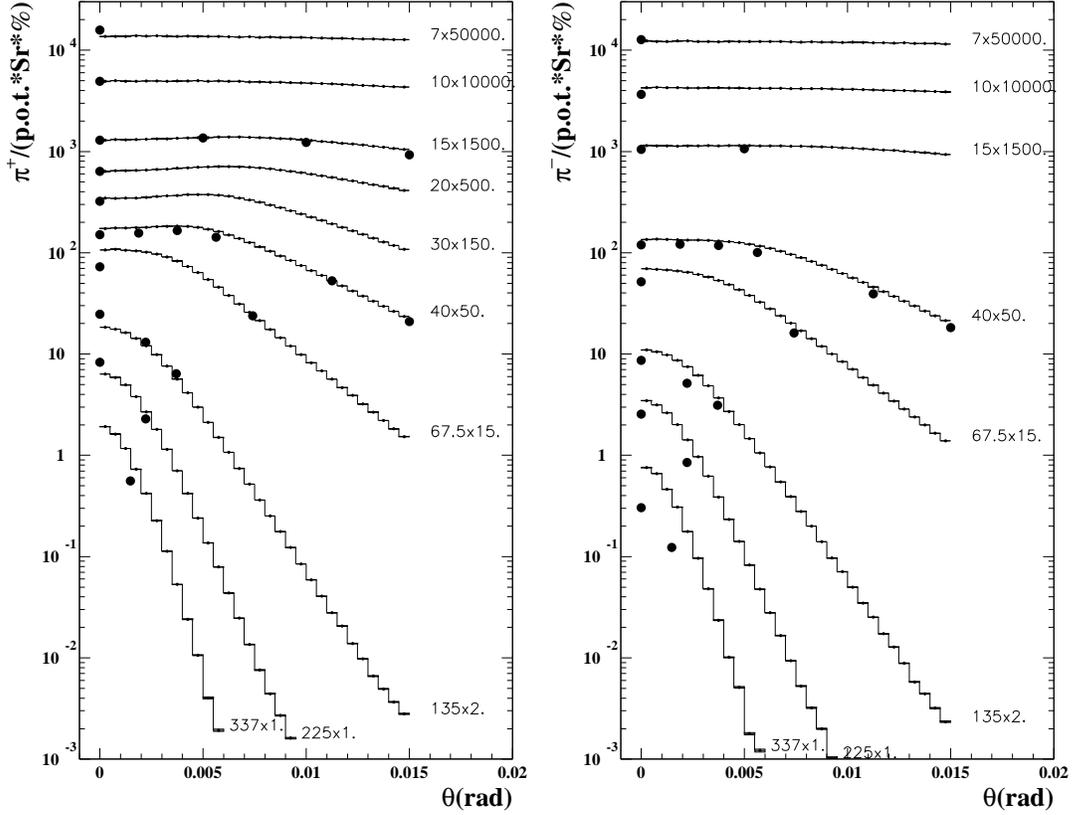}
\end{center}
\vspace*{-0.3cm}
\caption{Yields of $\pip$ (left) and $\pim$ (right) from p-Be
  interactions as a function of the production angle $\theta$ for
  different values of pion momentum. The predictions of FLUKA~2000 are
  shown as histograms, points represent the measurements of SPY
  and NA20.  The first number attached to each histogram is the
  $\pis$ momentum.  The second number is the factor by which both
  the data and the Monte Carlo prediction were rescaled to be
  accommodated on the plot.}
\label{fig:FlvsSPY}
\end{figure}

FLUKA~2000 was used to simulate
the yield of secondary particles from the interactions of 450~GeV
protons on the 100~mm thick Be target of SPY and NA20.
The results of these simulations were
compared to the two sets of particle production measurements described
in Section~\ref{sec:yields}. The comparison was carried out as a function
of the secondary particle momentum, $p$, and the production angle
$\theta$. It was
found~\cite{OPEN98} that the yields of secondary $\pis$ and $\ks$ agree
with the experimental data at the level of $\sim$20\% or better
with only a few exceptions, mostly for negative kaons or at large
momenta; the comparison plots, of which Fig.~\ref{fig:FlvsSPY} is
an example, can be found in Ref.~\cite{OPEN98}. This level of agreement
was considerably better than that
obtained with the generators of hadronic interactions implemented
within GEANT (such as GFLUKA or GHEISHA) and led us to choose
FLUKA for the simulation of the primary p-Be interactions.

However, in order to increase further the accuracy of the Monte Carlo
predictions, it was necessary to modify the FLUKA 2000 yields of secondary
particles in order to take into account the SPY and NA20 results.
The method used for this adjustment is described in the next section.

\subsection{Corrections for SPY and NA20 results}
\label{sec:reweight}
Corrections were introduced by ascribing a weight to each secondary
particle of a given type, $p$, and $\theta$, generated by FLUKA
in the p-Be interaction. Ideally, the value of the weight would
be obtained as the ratio between the measured rate and
the rate predicted by FLUKA for each particle
type and each $p$ and $\theta$. In practice the amount of available
data is limited, in particular for values of $\theta$
different from 0$^\circ$ (see Section~\ref{sec:yields}). The weights were
therefore calculated as a function of particle type and momentum only,
averaging over production angles whenever possible.  At 67.5 and
135~GeV/$c$ measurements were available from both NA20 and SPY.
At each momentum they were found to agree within the quoted errors and,
therefore, the average of the two, weighted according to the statistical
errors of the two measurements, was used. 

\begin{figure}
\vspace*{0.3cm}
\begin{center}
  \includegraphics*[scale=0.69]{\epsdir/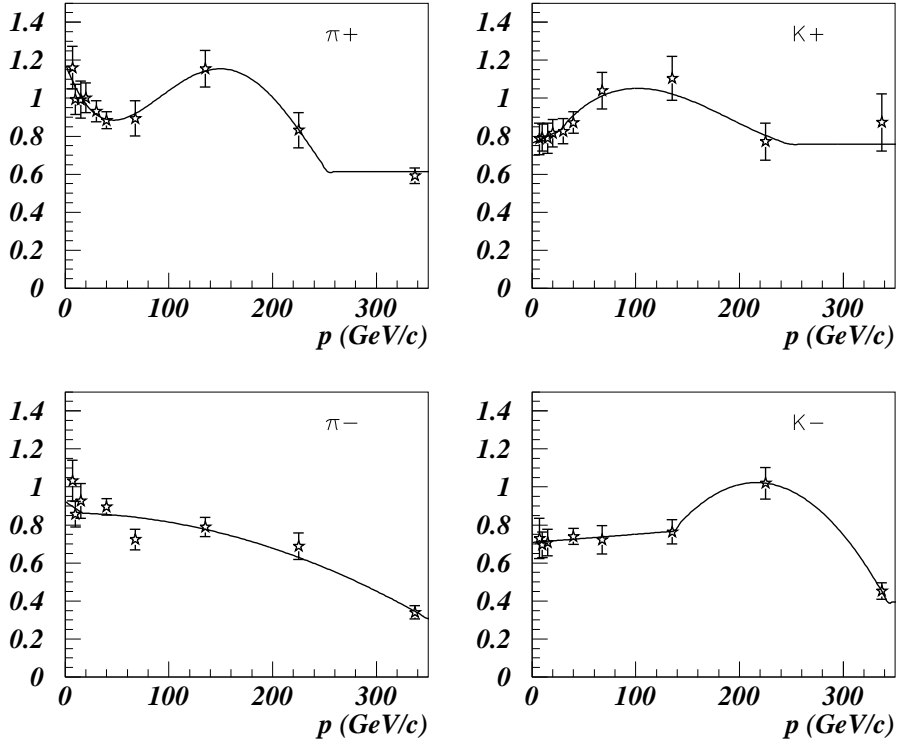}
\end{center}
\caption{The reweighting functions for $\pip$, $\kp$, $\pim$ and $\km$
  obtained from the SPY and NA20 measurements.  The points are the
  weights calculated for the values of $p$ for which the measurements
  were made, the curves are the result of fitting them with combinations
  of polynomial functions.}
\label{fig:fitExample}
\end{figure}

For those values of $p$ for which angular measurements were performed
(notably at 15 GeV/$c$ and 40 GeV/$c$), both the experimental and FLUKA
yields of secondary particles were convoluted with the WANF angular
acceptance functions and then integrated over all angles measured.
The ratio of these two integrals was defined as the weight at a given $p$. 
For the values of $p$ for which only the yields in the forward direction
($\theta = 0^\circ$) were measured (mostly below 40 GeV/$c$) the weights
were simply the ratios of the measured to predicted yields at $0^\circ$.
This is justified by the fact that at these low momenta the dependence
of the yield on the production angle is small below 10~mrad, the
acceptance of the beam line (Fig.~\ref{fig:FlvsSPY}).  These sets of
weights, obtained at discrete momenta, were fitted to combinations of
polynomial functions with systematic (see Section~\ref{sec:fluka_syst})
and statistical errors of the measurements combined in
quadrature and taken into account in the fits (see Fig.~\ref{fig:fitExample}
as an example). The resulting reweighting functions
were then applied on an event-by-event basis,
to every $\pis$, $\ks$, proton and antiproton emerging from the target
rod in which the primary interaction occurred.

The SPY experiment has measured the $K$/$\pi$ ratios with a much smaller
uncertainty than that of the separate measurements of the $K$
and $\pi$ yields, because of
partial cancelations of systematic uncertainties when taking
the ratio of yields measured under similar experimental conditions.
As can be seen in Table~\ref{tab:KoverPi}, the $K$/$\pi$ ratios
in our simulations agree with the SPY and NA20 results within
the quoted uncertainties.  This can be considered as an additional test
of the validity of our approach.

\begin{table}[thb]
  \caption{Comparison of the $K^+/\pi^+$ ratio predicted by our
    simulation, $(K^+/\pi^+)_{\rm MC}$\,, and the one measured in
    SPY and NA20, $(K^+/\pi^+)_{\rm data}$\,, for different values
    of the momentum $p$ and of the production angle $\theta$.}
  \begin{center}
    \begin{tabular}{cccc} \hline
      $p$ (GeV/$c$) & $\theta$ (mrad) & $(K^+/\pi^+)_{\rm MC}$ & $(K^+/\pi^+)_{\rm data}$ \\ \hline
      15.0      &  0.0       & $0.089 \pm 0.012$  & $0.083 \pm 0.003$ \\
                &  5.0       & $0.080 \pm 0.011$  & $0.081 \pm 0.003$ \\
                & 10.0       & $0.077 \pm 0.011$  & $0.081 \pm 0.002$ \\
      20.0      &  0.0       & $0.096 \pm 0.011$  & $0.097 \pm 0.002$ \\
      30.0      &  0.0       & $0.110 \pm 0.011$  & $0.106 \pm 0.002$ \\
      40.0      &  0.0       & $0.116 \pm 0.010$  & $0.110 \pm 0.002$ \\
                &  1.9       & $0.109 \pm 0.009$  & $0.104 \pm 0.002$ \\
                &  3.8       & $0.098 \pm 0.009$  & $0.092 \pm 0.002$ \\
                &  5.6       & $0.095 \pm 0.008$  & $0.092 \pm 0.002$ \\
                & 11.3       & $0.127 \pm 0.009$  & $0.132 \pm 0.002$ \\
      67.5      &  0.0       & $0.107 \pm 0.015$  & $0.105 \pm 0.001$ \\
                &  7.4       & $0.141 \pm 0.020$  & $0.140 \pm 0.003$ \\
      135.0     &  0.0       & $0.106 \pm 0.015$  & $0.081 \pm 0.001$ \\
                &  2.2       & $0.111 \pm 0.015$  & $0.118 \pm 0.002$ \\
                &  3.7       & $0.131 \pm 0.017$  & $0.154 \pm 0.002$ \\ \hline
    \end{tabular}
    \label{tab:KoverPi}
  \end{center}
\end{table}

Since no measurements of the $\klong$ and $\kshort$ yields are
available at these energies, they were estimated from the SPY
measurements of $\kp$ and $\km$ yields using the ``quark-counting''
method of Ref.~\cite{ND}.  This relation is:

\begin{equation}
\klong = \kshort = \frac{\kp+(2n-1)\km}{2n}\,, \label{eq:krel}
\end{equation}

where $n$ is the ratio of the u to d structure functions of the proton
evaluated at $x_R$, the ratio of the kaon energy in the centre of mass
to its maximum possible energy at its $p_T$.  These estimates were
then used to reweight the FLUKA 2000 yields.

Systematic uncertainties in the prediction of neutrino fluxes
arising from this reweighting procedure are discussed in
Section~\ref{sec:fluka_syst}.

\subsection{Transport and decays in the target region}
\label{sec:reint_target}
Transport of the secondaries within the boundaries of the target
box, including their possible decays and reinteractions in the target
rods downstream of the primary interaction vertex and in the box walls,
was handled by FLUKA 2000. The position and momentum vectors
of all the particles emerging from the target box and reaching the
upstream end of the copper collimator (115~cm from the centre
of the target) were saved to a file; their transport, reinteractions
and decays in the beam line downstream of the target were performed
later by NUBEAM in a separate simulation run.

A small fraction of the overall neutrino flux is produced directly
in the target region. It comes primarily from the prompt decays of
charmed mesons as well as $\pis$ and $\ks$ decays (see
Section~\ref{sec:decays}).

\subsection{Magnetic field}
\label{sec:B}
 An accurate description of the magnetic field in the
 horn and the reflector is extremely important for the
 prediction of both the major component of the beam, $\numu$, and
 its minor components, $\anumu$, $\nue$ and $\anue$.

\begin{figure}
\begin{center}
  \includegraphics*[scale=0.50]{\epsdir/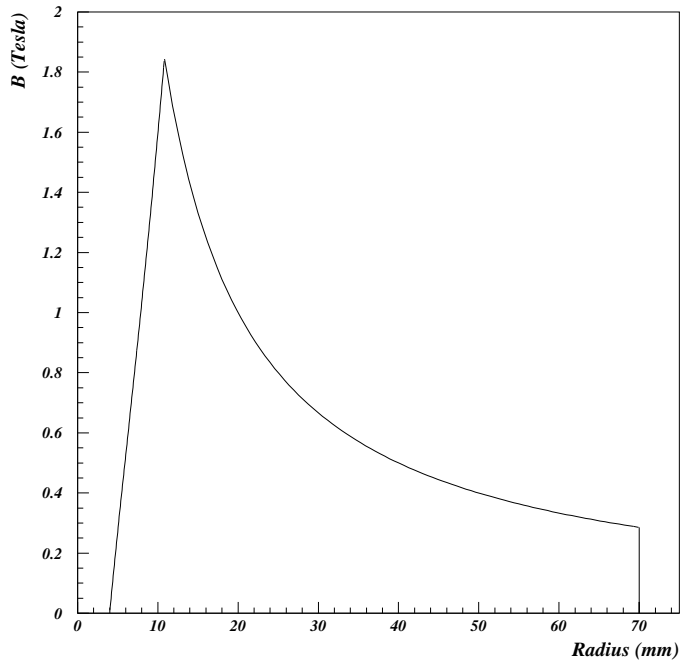}
\end{center}
\vspace*{-0.3cm}
\caption{Magnetic field in the horn as a function of its radius, at the
  downstream end (the neck) of the horn. The radial position of the inner
  conductor is between 4~mm and 10.8~mm.}
\label{fig:bhorn}
\end{figure}

 The magnetic field in the region between the two coaxial inner and
 outer conductors was simulated according to Eq.~(\ref{eq:bfield}).
 The measurements of the magnetic field in a spare horn (identical
 to the one installed in the WANF) revealed no deviations from
 the expected behavior. The magnetic field in the inner conductors
 of the horn and reflector was also taken into account, with
 the current skin depth calculated using the Fourier transform
 of the horn pulses. The radial dependence of the magnetic field
 in the horn is shown in Fig.~\ref{fig:bhorn}. The maximum value
 of the field (1.85~T for the horn and 0.24~T for the reflector)
 occurs at the downstream extremities of both focusing elements,
 at the outer surface of the inner conductor.

 Transport of the particles in the magnetic field was performed with
 the Runge-Kutta method; special care was taken in optimizing the appropriate
 GEANT tracking medium parameters.

\begin{figure}
\begin{center}
  \includegraphics*[scale=0.65]{\epsdir/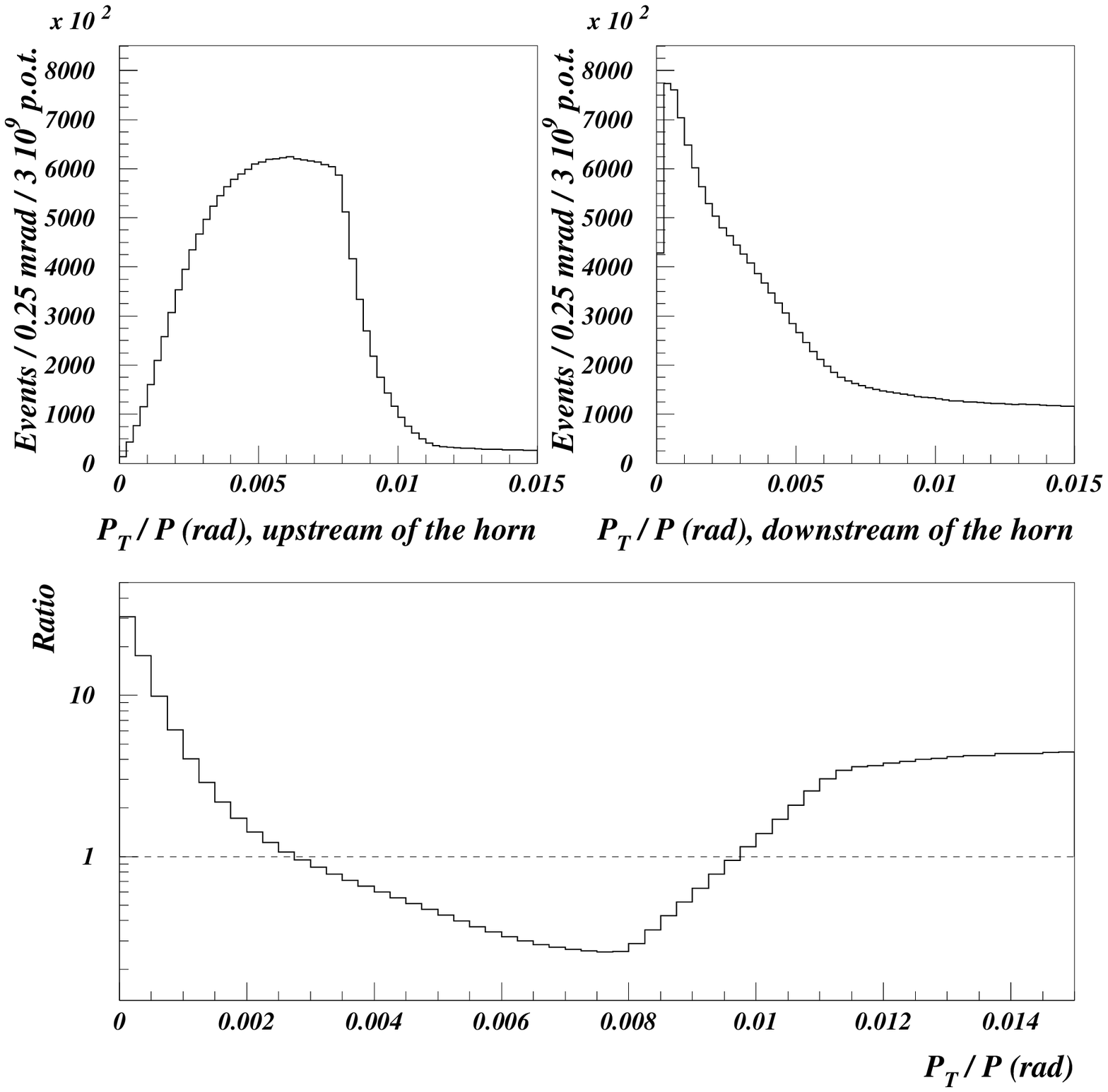}
\end{center}
\vspace*{-0.4cm}
\caption{Distribution of the angle between the $\pip$ momentum vector
  and the beam line direction, $p_T/p$, just upstream of the horn (top
  left), right after it (top right) and the ratio of the latter to
  the former (bottom).}
\label{fig:horn_pi+}
\end{figure}

\begin{figure}
\begin{center}
  \includegraphics*[scale=0.65]{\epsdir/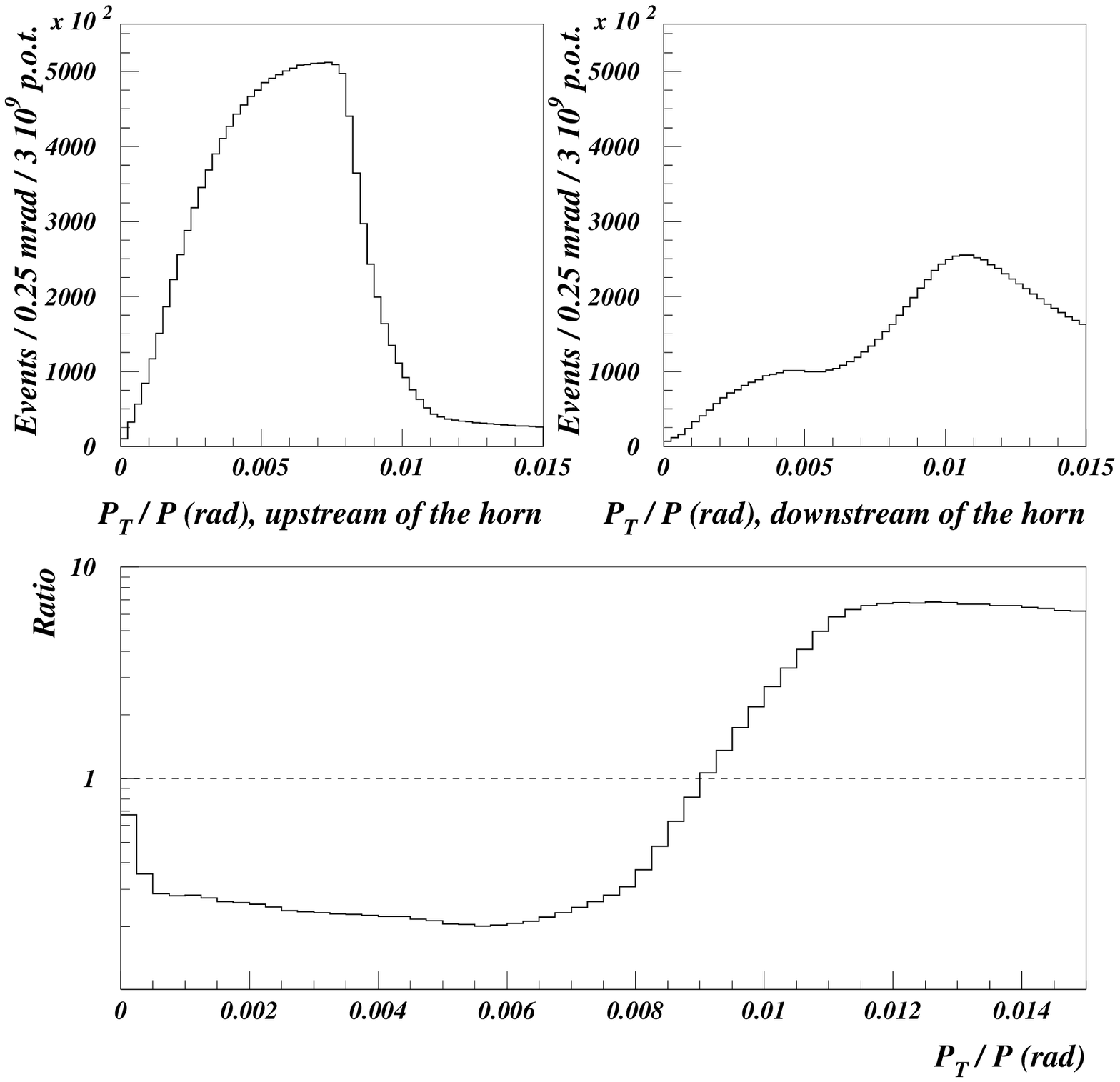}
\end{center}
\vspace*{-0.4cm}
\caption{Distribution of the angle between the $\pim$ momentum vector
  and the beam line direction, $p_T/p$, just upstream of the horn (top
  left), right after it (top right) and the ratio of the latter to
  the former (bottom).}
\label{fig:horn_pi-}
\end{figure}

 The effect of the horn and of the reflector on particles of different
 signs is illustrated in Figs.~\ref{fig:horn_pi+} and \ref{fig:horn_pi-},
 which show angular distributions of positive and negative pions at a
 plane just upstream of the horn and immediately downstream
 of it. Upstream of the horn, pions of both charges emerging from the
 target have very similar angular distributions, with the bulk of the
 particles within $\sim$10~mrad, which is the acceptance of the
 collimators. While traversing the horn, positive pions with momentum
 around 50~GeV/$c$ are focused into a near-parallel beam leading to an overall
 enhancement at small angles of up to a factor of 30 (Fig.~\ref{fig:horn_pi+}).
 Negative pions are strongly defocused resulting in their reduction at
 small angles by as much as a factor of 5 (Fig.~\ref{fig:horn_pi-}).
 The reflector provides an additional focusing for
 positive particles of momentum both higher and lower than 50~GeV/$c$
 that were respectively underfocused and overfocused by the horn.

\begin{figure}
\begin{center}
  \includegraphics*[scale=0.65]{\epsdir/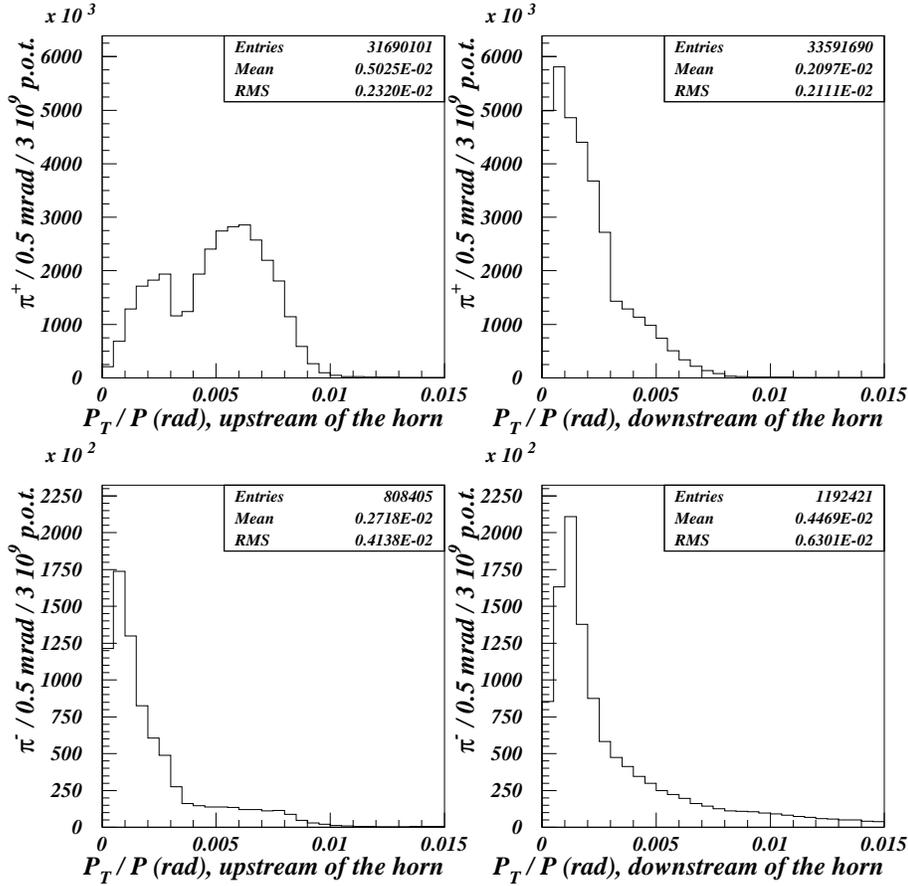}
\end{center}
\vspace*{-0.5cm}
\caption{Distribution of the angle between the pion momentum vector
  and the beam line direction, $p_T/p$, just upstream of the horn (left)
  and immediately after it (right), for $\pip$ producing a $\numu$
  at NOMAD (top) and $\pim$ producing a $\anumu$ (bottom).}
\label{fig:horn_pi_nu}
\end{figure}

 Fig.~\ref{fig:horn_pi_nu} shows again angular distributions of $\pip$ and
 $\pim$ upstream and downstream of the horn but now only those $\pip$ that
 ultimately produce a $\numu$ reaching the NOMAD detector (and $\pim$
 that give a $\anumu$) are included. From the left-hand plots it can be
 seen that only mesons produced with angles smaller than $\sim 10$~mrad
 can produce neutrinos that traverse the NOMAD detector. The distribution
 of $\pip$ upstream
 of the horn has two distinct regions: the first, at small angles,
 is mainly populated by high-energy (and hence very forward) pions;
 the second, at larger angles, -- by low- and medium-energy (up to about
 100~GeV) pions. The focusing effect of the horn on $\pip$ (and $\kp$)
 in the first region is modest; however, it is crucial for the particles
 in the second region: their deflection is such (Fig.~\ref{fig:horn_pi_nu},
 top right) that they enter the decay tunnel and contribute to the $\numu$
 flux at NOMAD. The minor component, $\anumu$, of the neutrino flux
 comes from the decays of very forward $\pim$ and $\km$ that could
 not be defocused (Fig.~\ref{fig:horn_pi_nu}, bottom).  Another
 important source of $\anumu$'s are interactions in the horn and in
 material further downstream, which are discussed in the next section.

 Overall, the WANF horn--reflector system provides more than a factor
 of four increase in the $\numu$ flux at the NOMAD detector in the relevant
 energy range (between 2 and 200 GeV); the admixture of $\anumu$ in
 the beam is at the same time reduced from 70\% to less than 7\%. The reflector
 provides an increase of about 25\% in the $\numu$ flux compared with
 the horn-only case.

\subsection{Secondary interactions}
\label{sec:mat}
 Reinteractions of secondary particles in the beam line hardware elements
 situated downstream of the target affect the neutrino fluxes. Their impact
 on antineutrino components of the beam is particularly large since the mesons
 of the ``wrong'' sign produced in the secondary interactions downstream
 of the focusing elements are not defocused and some of them then decay
 in the decay tunnel. A fraction of primary protons, which either
 did not interact in the target or missed it geometrically, also interacts
 in the material downstream of the target resulting in yet another contribution
 to neutrino fluxes. Therefore, an accurate description of the beam line
 hardware and of the particle yields from interactions downstream of
 the target is essential.

 A detailed simulation of all the elements constituting the WANF
 beam line (described in Section~\ref{sec:beamhard}) was performed.
 The NUBEAM description of the horn and the reflector included the
 insulating spacers, flanges, conductor plates and water cooling
 hardware. Special attention was paid to the simulation of the downstream
 (and closest to the beam axis) part of the horn inner conductor, the
 neck, since it is traversed by the large flux of very forward particles,
 including primary protons. Cables, screws, bolts and nuts were
 approximated by disks of iron of appropriate thickness. The titanium
 windows of the helium bags, as well as the surrounding pipe and flanges,
 were included. All measuring devices installed in the line
 (ionization chambers and SEM's) were also simulated. Finally,
 the entrance window of the decay tunnel, a system of support rings,
 the tunnel walls and the hardware contained in the tunnel were
 also included in the simulation.

 In the GEANT3 framework, the most suitable model for the simulation
 of particle yields from the secondary hadronic interactions is GFLUKA.
 However, the FLUKA package has undergone significant
 improvements~\cite{GFL_FLS}
 since the time when its 1992 version was implemented into GEANT.
 In NUBEAM, these improvements were taken into account by correcting
 the GFLUKA meson yields by the corresponding differences between FLUKA
 2000 and GFLUKA.  Since it was not possible
 to implement FLUKA 2000 in GEANT3 and therefore in NUBEAM,
 special simulation runs were performed in which the beam line material
 downstream of the target was replaced by a 0.5 interaction length slab
 of aluminium placed at the position of the neck of the horn.  The thickness
 of the slab corresponded to the average amount of material traversed
 by secondary particles contributing neutrinos at NOMAD and the position
 of the slab to the most likely reinteraction point along the beam line.
 Two such special runs were performed, one with GFLUKA in GEANT3 and
 one with FLUKA 2000, both runs using as input the same set of
 particles, which had the composition, momentum and angular distributions
 of the ones produced in the simulation of p-Be interactions.

 The comparison showed that the yields of tertiary $\pip$ and
 $\pim$ in FLUKA 2000 were smaller by about 30\% at all
 energies, whereas those of $\kp$, $\km$ and $\klong$ were
 larger by up to 30\% at energies below 30~GeV; these differences
 had only a very weak dependence on the production angle.  The
 corrections obtained were applied on an event-by-event basis
 in the standard NUBEAM runs,
 as energy-dependent weights to $\pi$'s and $K$'s produced
 in secondary interactions.  Their net effect was a reduction of
 about 10\% in the $\anumu$ flux at energies below 15~GeV, and
 an increase of $\sim 5$\% in the $\nue$ and $\anue$ fluxes (and in the
 $\emurat$ ratio) in the same low energy region.



\begin{figure}
\begin{center}
  \includegraphics*[scale=0.65]{\epsdir/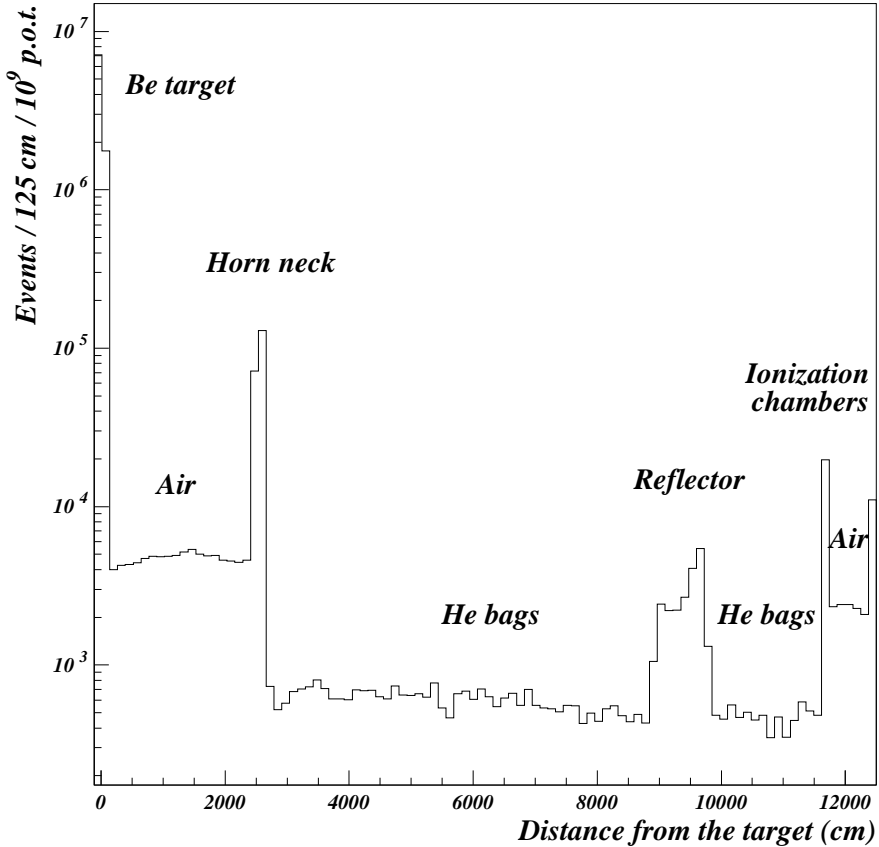}
\end{center}
\vspace*{-0.4cm}
\caption{Longitudinal position of proton interactions resulting in a neutrino
 (of any flavour) at NOMAD. The position is measured relative to the
 centre of the target; the horizontal scale extends up to the entrance
 window of the decay tunnel. The locations of the main beam elements
 are also indicated.}
\label{fig:prot_int}
\end{figure}

 The effect of the interactions of the incident protons, which either did
 not interact in the target or missed it geometrically, in the hardware
 downstream of the target on the neutrino flux at NOMAD is illustrated in
 Fig.~\ref{fig:prot_int}, which shows the NUBEAM prediction of the
 position of the interaction of the primary protons along the beam line;
 only events resulting in a neutrino reaching the NOMAD detector are
 included. About
 3\% of both the $\numu$'s and the $\nue$'s at NOMAD were found to
 originate from proton interactions downstream of the target -- mainly
 in the narrowest part of the horn.  The corresponding contribution
 is larger for the $\anumu$ flux: about 15\% of the $\anumu$'s
 result from proton interactions downstream of the target. The
 reason is that the negatively charged mesons produced in the
 neck of the horn or further downstream are not (or only weakly)
 defocused and have a larger probability of entering the decay
 tunnel compared with the ``wrong'' sign mesons produced in the
 Be target.

\begin{figure}
\begin{center}
  \includegraphics*[scale=0.65]{\epsdir/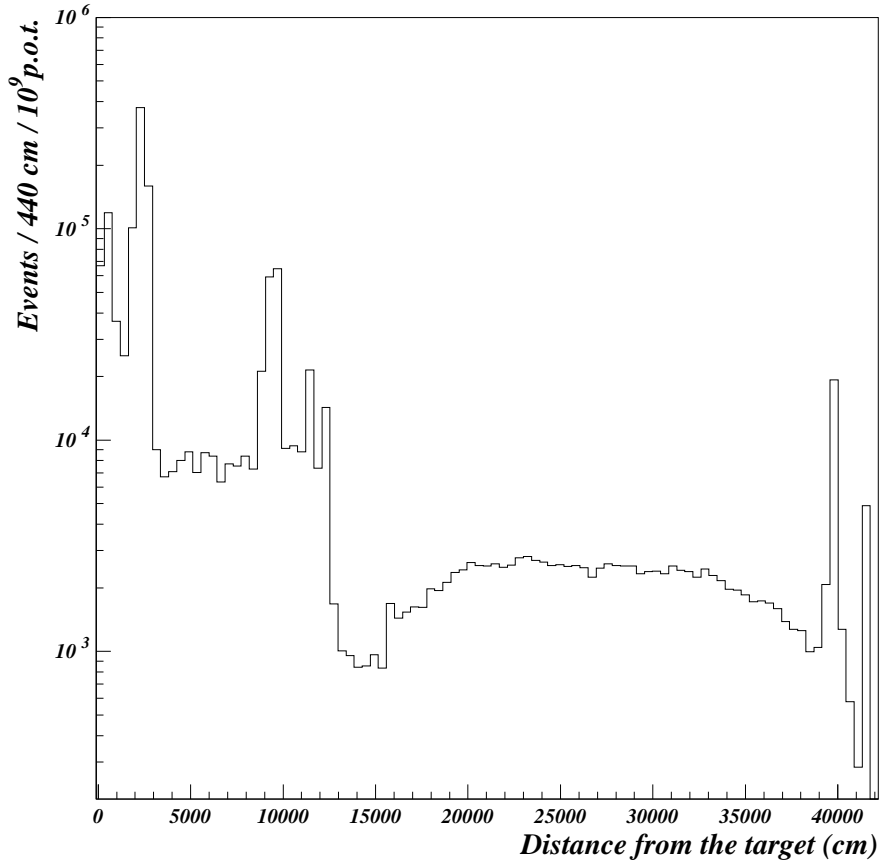}
\end{center}
\vspace*{-0.4cm}
\caption{Longitudinal position of secondary interactions resulting in
 a neutrino (of any flavour) at NOMAD.
 The horizontal scale extends up to the iron filter.}
\label{fig:sec_int}
\end{figure}

 Since the ``parents'' of neutrinos reaching
 NOMAD traverse an amount of material equivalent, on average, to about
 half an interaction length, reinteractions play
 an important role in the production of the neutrino beam. Secondary
 interactions resulting in neutrinos reaching NOMAD occur mainly
 in the horn, the collimators and the reflector (Fig.~\ref{fig:sec_int}).
 Overall, secondary interactions
 in the material downstream of the target produce about 10\% of
 $\numu$'s, 12\% of $\nue$'s and 45\% of $\anumu$'s at NOMAD.
 The energy spectrum of these neutrinos is significantly softer than
 the one of neutrinos produced in the decays of mesons which did not
 experience secondary interactions: the average energies of these two
 components of the flux are, respectively, 16.7 GeV and 25.2 GeV for
 $\numu$ and 19.4 GeV and 39.6 GeV for $\nue$. Since a \nmne oscillation
 signal would manifest itself as an excess of $\nue$ events at low
 energies, an accurate description of the material downstream of the
 target was crucial to the \nmne oscillation search.

\subsection{Decays}
\label{sec:decays}
 High energy neutrinos are produced in two- and three-body decays in flight
 of mainly $\pis$, $\ks$, $\klong$, $\mus$ and charmed mesons. In the
 default GEANT3, all decays are treated according to pure phase-space.
 We modified the GEANT version used in NUBEAM to treat semileptonic
 $\ks$ and $\klong$ decays taking into account the V--A structure
 of the weak leptonic current and the $K_{e3}$ form factors~\cite{PDG}.
 Pure V--A muon decays were simulated assuming that the muon (produced
 mainly in pion decays) is fully polarized. Charmed mesons and strange
 baryons
 were added to the GEANT particle list, with their relevant decay
 modes and branching ratios defined according to Ref.~\cite{PDG}.
The charmed particles were mainly produced at the beryllium target and
at the hadron filter; their production cross section in p-Be interactions
was taken to be 0.45~mb~\cite{FLUKA}.
The contribution from this source
to neutrinos at NOMAD was small: 3.2\% for $\anue$, 0.6\% for
$\nue$, 0.1\% for $\anumu$ and negligible for $\numu$.

 In order to generate a sufficient number of neutrino events in a reasonable
 time, the decay
 of each particle with a neutrino among its decay products was repeated
 100 times; each time the decay mode was randomly chosen according to
 the branching ratios and the kinematics of the decay generated anew.
 It was shown that this procedure does not lead to any significant
 bias once the total number of generated events is large; the effect
 of this procedure on statistical errors of Monte Carlo distributions
 was also studied and appropriately taken into account.

\section{Beam composition}
\label{sec:fluka_comp}
 The spectra of the four principal neutrino species, $\numu$, $\anumu$,
 $\nue$ and $\anue$\,, and of their components, predicted by the simulation
 described above, are shown in Fig.~\ref{fig:fluka_comp}. The average
 energies and the relative abundances of the four neutrino species,
 as well as the relative contributions to the neutrino fluxes from
 $\pis$, $\ks$ and other sources and their average energies, are listed
 in Table~\ref{tab:fluka_comp}. We can summarize them as follows:

\begin{figure}
\vspace*{-0.3cm}
\begin{center}
  \includegraphics*[scale=0.69]{\epsdir/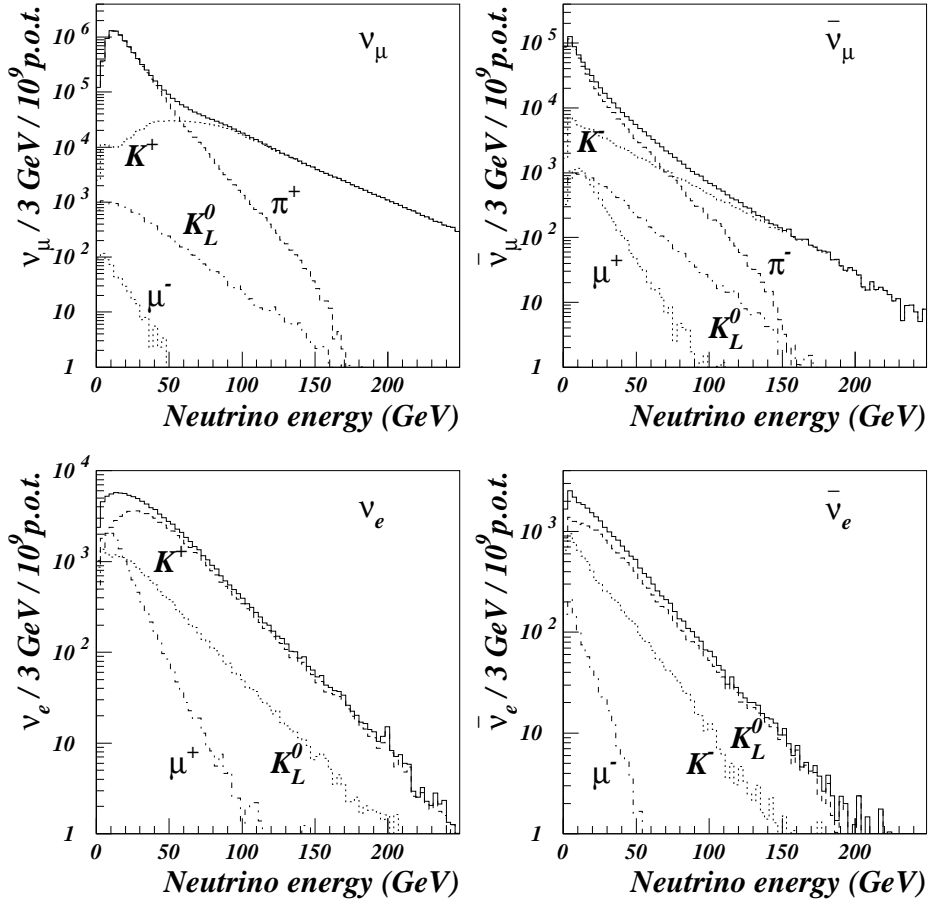}
\end{center}
\vspace*{-0.4cm}
\caption{Composition of the $\numu$, $\anumu$, $\nue$ and
  $\anue$ energy spectra at NOMAD, within the transverse fiducial area
  of 260$\times$260~cm$^2$.}
\label{fig:fluka_comp}
\end{figure}

\begin{table}[thb]
 \caption{Composition of the neutrino beam and its various species.}
 \begin{center}
 \begin{tabular}{cccrrrrrrrr} \hline
 \multicolumn{1}{c}{} & \multicolumn{2}{c}{} &
 \multicolumn{8}{c}{Source}\\ \cline{4-11}
 \multicolumn{1}{c}{$\nu$} & \multicolumn{2}{c}{Flux} &
 \multicolumn{2}{c}{$\pip$ or $\pim$} & \multicolumn{2}{c}{$\kp$ or $\km$} &
 \multicolumn{2}{c}{$\klong$} & \multicolumn{2}{c}{$\mup$ or $\mum$}\\
 \cline{2-11}
 species &  Abund.  & $\langle E_\nu \rangle$ &
         \% & $\langle E_\nu \rangle$ & \% & $\langle E_\nu \rangle$ &
         \% & $\langle E_\nu \rangle$ & \% & $\langle E_\nu \rangle$ \\ \hline
 $\numu$  &  1.0    & 24.3 &
        90.4 & 19.1 \ &  9.5 & 73.0 \ &  0.1 & 26.8 \ & $<$0.1 & 11.4 \\ \hline
 $\anumu$ &  0.0678 & 17.2 &
        84.0 & 13.8 \ & 12.8 & 38.1 \ &  1.9 & 26.9 \ &    1.2 & 17.0 \\ \hline
 $\nue$   &  0.0102 & 36.4 &
         --  &  --  \ & 68.0 & 41.8 \ & 17.8 & 30.3 \ &   13.6 & 16.8 \\ \hline
 $\anue$  &  0.0027 & 27.6 &
         --  &  --  \ & 25.1 & 22.8 \ & 68.2 & 30.4 \ &    3.5 & 11.1 \\ \hline
 \end{tabular}
 \end{center}
\label{tab:fluka_comp}
\end{table}

\begin{itemize}
 \item The $\numu$ neutrinos are primarily produced via two-body decays
    of $\pip$ (90.4\% of $\numu$) and $\kp$ (9.5\%), with much smaller
    contributions from other sources ($\klong$, $\mum$, charmed hadrons,
    etc.). Neutrinos from pion decays dominate the $\numu$ spectrum
    up to $\sim 60$~GeV, whereas those from $\kp$ decays dominate
    beyond this energy.
 \item Similar to $\numu$, the $\anumu$ neutrinos are primarily
    produced via decays of $\pim$ (84.0\% of $\anumu$) and $\km$
    (12.8\%). Compared to $\numu$, a larger fraction of $\anumu$
    comes from $\klong$, $\mup$, and charmed hadron decays since
    these particles are not affected by the defocusing of the horn and
    reflector. The $\km/\pim$ ratio being smaller than the $\kp/\pip$
    ratio, the $\anumu$ from $\km$ only start to dominate the $\anumu$
    spectrum at about 70~GeV. The $\anumu$ flux is 6.8\% that of the $\numu$.
 \item Four decays contribute to the $\nue$ flux. The main contribution
    is from $\kp \ra$ $\pi^0$ $e^+$ $\nue$ (68.0\% of $\nue$),
    followed by the $\klong$ $\ra$ $\pim$~$e^+$~$\nue$ (17.8\%),
    muon decays (13.6\%) and charmed hadron and hyperon decays (0.6\%).
    The $\nue$ flux relative to $\numu$ in the absence of \nmne
    oscillations is expected to be about 1.0\% when integrated over
    all energies and 0.5\% below 20 GeV.
 \item The principal source of $\anue$ is the
    $\klong$ $\ra$ $\pip$~$e^-$~$\anue$ decay, accounting for about
    68\% of $\anue$. The other sources of $\anue$ are:
    $\km$ $\ra$ $\pi^0$~$e^-$~$\anue$ ($\approx 25$\%), charmed hadron
    decays (3.2\%), and a small contribution from $\mum$.
    The $\anue$ flux relative to $\numu$ is about 0.3\%.
\end{itemize}

\section{Systematic uncertainties}
\label{sec:fluka_syst}
As explained in Section~\ref{sec:mc}, neutrinos in the beam originate from
the decay of mesons produced through four different mechanisms: proton-Be
interactions in the target, proton interactions downstream of the target
in material other than beryllium, reinteractions of particles in the target
and reinteractions of particles downstream of the target.  The systematic
uncertainties on the yields of particles from proton-Be interactions in
the target are described in Section~\ref{sec:syst_target}.  The
uncertainties arising from the yields of particles in interactions other
than proton-Be are described
in Section~\ref{sec:syst_reint}.  Other sources of systematic uncertainties,
such as those arising from the position of the beam relative to the target
and from the propagation of secondary mesons through the WANF beam line,
are described in Section~\ref{sec:syst_others}.  The summary of uncertainties
is given in Section~\ref{sec:syst_summary}.

\subsection{Uncertainty on the yields of particles from p-Be interactions}
\label{sec:syst_target}
The main source of the systematic uncertainties in the prediction
of the $\emurat$ ratio was due to the uncertainty on the yields
of secondary particles from p-Be interactions.  This 
was estimated in two steps.  First the overall relative systematic
uncertainty, $\Delta$, on the reweighting function of each particle
type yielding neutrinos was
estimated.  This uncertainty was particle type and momentum dependent.
Then, the effect of $\Delta$ on each neutrino species and on the
$\emurat$ ratio was computed.

The first step in the calculation of $\Delta$ was to identify the
systematic uncertainties of SPY and NA20 that did not cancel in the
$\emurat$ ratio, namely those due to the particle selection efficiency and
identification, the particle-dependent losses along the spectrometer,
the particle decays, and the stability of the intensity of the proton
beam and of its position relative to the target.  The ones that did
cancel amounted to 1.8\% and were removed from the published SPY and NA20
systematic uncertainties.  The systematic uncertainties were
combined in quadrature with the statistical uncertainties of the measurements.
The relative error on the reweighting function arising from this source
is referred to as $\Delta_1$.

The uncertainty arising from using a single, angle-independent,
correction for each momentum and particle type, $\Delta_2$, was estimated
as follows.  For those values of momentum $p$ for which angular scans were
available, the uncertainty was estimated as the root mean square
deviation between the individual angular measurements available at that
$p$ and the results of a single-valued reweighting obtained from these
measurements as described in Section~\ref{sec:reweight}.  For values
of $p$ at which only the 0$^\circ$ angle was measured, mostly below
40 GeV/$c$, the uncertainty was taken
to be the difference between the 0$^\circ$ point and the single-valued
reweighting at the angular scan nearest in momentum.  This is justified
by the fact that, as was noted in Section~\ref{sec:reweight}, at these
low momenta the dependence of the yield
on the production angle is small below 10~mrad. 
For each momentum and particle type, $\Delta_1$ and $\Delta_2$ were
combined in quadrature, to give $\Delta$, the error used in the
fits described in Section~\ref{sec:reweight}.  As an example
the values of $\Delta_1$, $\Delta_2$ and $\Delta$ for $\pip$ are
given in Table~\ref{tab:pi_error}.

\begin{table*}
  \caption{The values of $\Delta_1$, $\Delta_2$ and $\Delta$ (described
    in the text) for $\pip$, at different values of the momentum $p$.}
  \begin{center}
    \begin{tabular}{l|ccccccccc} \hline
      $p$ (GeV/$c$)           &   7   &   10  &  15   &  20   &  30   &  40   & 67.5  &  135  &  225  \\ \hline
      $(\Delta_1)_{\rm stat}$ & 0.010 & 0.010 & 0.003 & 0.003 & 0.005 & 0.003 & 0.004 & 0.004 & 0.005 \\
      $(\Delta_1)_{\rm syst}$ & 0.096 & 0.079 & 0.098 & 0.078 & 0.057 & 0.046 & 0.051 & 0.062 & 0.065 \\ 
      $\Delta_2$              & 0.007 & 0.007 & 0.004 & 0.007 & 0.014 & 0.020 & 0.089 & 0.056 & 0.091 \\
      $\Delta$                & 0.097 & 0.080 & 0.098 & 0.078 & 0.059 & 0.050 & 0.103 & 0.084 & 0.112 \\ \hline
    \end{tabular}
  \end{center}
  \label{tab:pi_error}
\end{table*}

\begin{figure}
  \begin{center}
    \includegraphics*[scale=0.66]{\epsdir/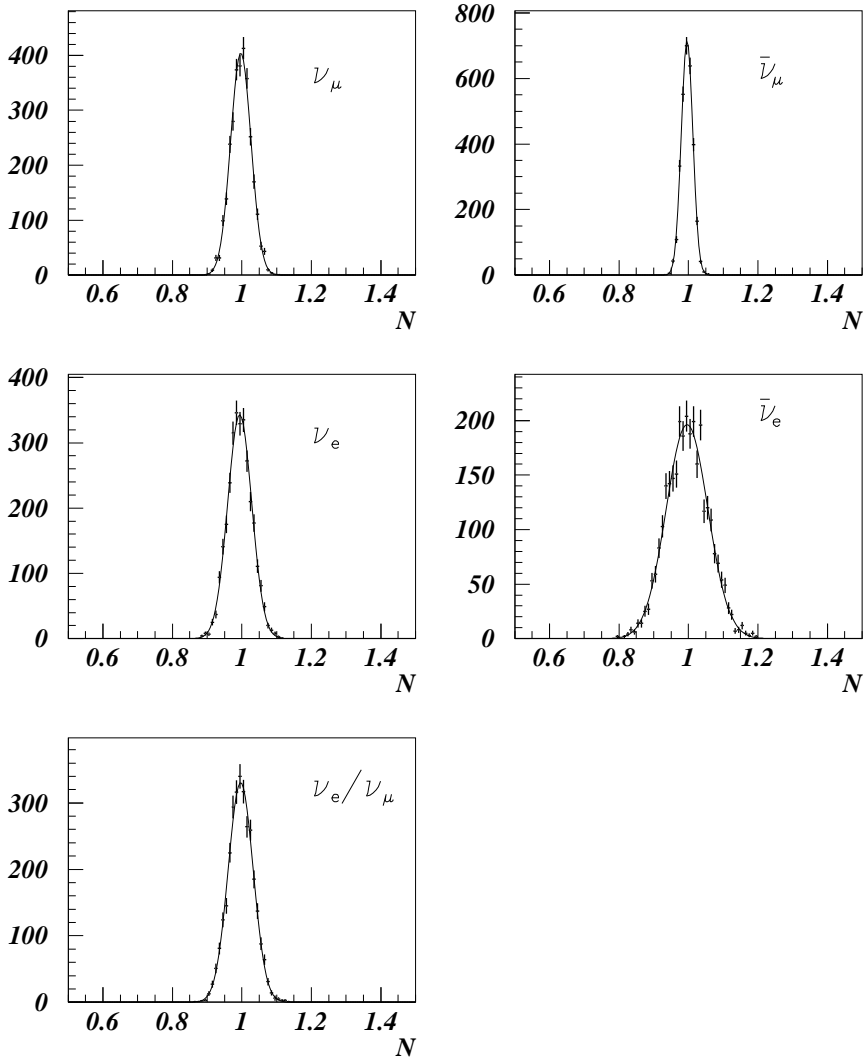}
  \end{center}
  \caption{Distributions of $N$ (see text) for each of the four neutrino
    species and for the $\emurat$ ratio.}
  \label{fig:err_norm_nus}
\end{figure}

The systematic uncertainty on the neutrino flux predictions at NOMAD
arising from $\Delta$ and from the use of a fit to interpolate between
the discrete experimental measurements of SPY and NA20 was then evaluated
as follows:
\begin{itemize}
\item The distributions of neutrinos at NOMAD were generated using NUBEAM
  and the values of the reweighting functions obtained using the fits as
  described in Section~\ref{sec:reweight}.
\item These neutrinos were separated into classes defined by the species
  of the neutrino and by the type of its parent particle emerging from the
  target rod in which the primary interaction occurred. For each such class
  a fine-binned two-dimensional histogram of the parent particle momentum
  versus the neutrino energy was filled.
\item Three thousand ``simulated experiments'' were then performed.  Each
  such experiment consisted of the following steps:
\begin{itemize}
\item The discrete values of the weights
  were modified at random about their central values according to Gaussians
  with $\Delta$ as standard deviations and the fits repeated.  This resulted
  in a new particle production prediction for $\pis$, $\ks$, protons and
  antiprotons.
\item The normalizations of the $\kp$ and $\km$ fits were further modified
  at random according to a Gaussian of 1.2\% width to take into account
  the uncertainty in the $K_{e3}$ branching ratio~\cite{PDG}.
\item The $\klong$ and $\kshort$ predictions were also recalculated by
  using the new $\kp$ and $\km$ predictions into the quark-counting
  formula~(\ref{eq:krel}) together with an additional uncertainty generated
  at random according to a Gaussian
  with a standard deviation of 15\% (the uncertainty in the accuracy
  of the formula).
\item The content of each bin of the two-dimensional class histograms
  was modified by the ratio of the new fit to the central fit evaluated
  for the parent particle type and at the momentum of that bin.
\item By summing the class histogram bins over the parent particle momentum
  and type, a new energy-dependent flux prediction at NOMAD was obtained
  for each neutrino species.
\item In order to separate the uncertainty into an energy-dependent
  uncertainty and a normalization uncertainty, the new integral
  flux of each neutrino species was compared to the integral flux obtained
  with the central fit.  The ratio of these two integrals, $N$, was used to
  renormalize the integral flux of the simulated experiment to
  that of the central fit.
\item Finally, the energy-dependent prediction of the $\emurat$ flux ratio
  was also obtained, as well as the ratio, $N_{e \mu}$,
  of the $\nue$ and $\numu$ normalization factors.
\end{itemize}
\item Repeating the simulated experiment 3000 times resulted in:
\begin{enumerate}
\item an envelope
  of predictions for each neutrino species and for the $\emurat$ ratio from
  which the energy-dependent systematic uncertainty, $\Delta_{\nu}$,
  was extracted. At any energy it was taken as the r.m.s. width of the
  envelope at that energy.  Representative
  values of $\Delta_{\nu}$ are listed in Table~\ref{tab:error_shape_T9}.
\item the distributions of the 3000 values of $N$ for each neutrino species
  and for $N_{e \mu}$.  They are shown in Fig.~\ref{fig:err_norm_nus}. Their
  standard deviations, 0.029 for $\numu$, 0.017 for $\anumu$, 0.035 for
  $\nue$, 0.060 for $\anue$ and 0.036 for $\emurat$, were used as the
  normalization uncertainty for each
  species and for the $\emurat$ ratio.  Note that due to correlations
  between the origins of $\numu$ and $\nue$ fluxes, the uncertainty on
  the $\emurat$ ratio is smaller than would be expected from the
  uncertainties on the individual $\numu$ and $\nue$ fluxes.   The standard
  deviation is smaller for $\anumu$ than for $\numu$ because this uncertainty
  only refers to $\numu$ and $\anumu$ originating from mesons produced
  directly in the target
  and, as explained in Section~\ref{sec:mat}, the fraction of $\anumu$
  at NOMAD from this source is smaller than the corresponding one from
  $\numu$. 
\end{enumerate}
\end{itemize}

\begin{table}
  \caption{Representative values of the energy-dependent systematic uncertainty
    $\Delta_{\nu}$ at selected values of neutrino energy $E$, for each of the
    four neutrino species and for the $\emurat$ ratio.}
  \begin{center}
    \begin{tabular}{c|cccccc} \hline
      $E$ (GeV)        &  10   &  30   &  50   &  70   &  100  &  130 \\ \hline
      $\numu$          & 0.017 & 0.016 & 0.028 & 0.038 & 0.055 & 0.065\\
      $\anumu$         & 0.007 & 0.007 & 0.015 & 0.018 & 0.027 & 0.036\\
      $\nue$           & 0.021 & 0.011 & 0.016 & 0.025 & 0.050 & 0.080\\
      $\anue$          & 0.030 & 0.011 & 0.022 & 0.031 & 0.040 & 0.055\\
      $\emurat$        & 0.025 & 0.020 & 0.030 & 0.038 & 0.057 & 0.065\\ \hline
    \end{tabular}
  \end{center}
  \label{tab:error_shape_T9}
\end{table}

The small contribution to the systematic uncertainty arising from the
yields of particles other than pions, kaons, protons and antiprotons
(referred to as ``others'') -- which included, among other things, the
contribution from a conservative uncertainty of 50\% assigned to the
production cross section of charmed mesons -- was also subdivided into
a normalization and an energy-dependent component.  They were combined
in quadrature with the standard deviations of $N$ (and $N_{e \mu}$)
and with $\Delta_{\nu}$, respectively.  Finally, the common
systematic uncertainty of 1.8\% that had been removed from the SPY and
NA20 results was recombined in quadrature with the normalization
uncertainties of the individual neutrino flavours resulting in the
normalization uncertainties from the yields of secondary particles from
the beryllium target shown in the first line of Table~\ref{tab:normerrors}.

\begin{table}[htb]
  \caption{Summary of energy-independent relative systematic uncertainties
    in the $\numu$, $\anumu$, $\nue$ and $\anue$ fluxes and in the
    $\emurat$ ratio.  The energy-dependent uncertainties are shown in
    Fig.~\ref{fig:err_nus}.}
  \begin{center}
    \begin{tabular}{l|ccccc} \hline
                                        & $\numu$ & $\anumu$ & $\nue$ & $\anue$ & $\emurat$ \\
      Source of uncertainty             &         &          &        &         &           \\ \hline
      Yields of secondary particles     &  0.034  &   0.029  &  0.039 &  0.064  &  0.036    \\
      Proton interaction downstream of target & 0.002 & 0.024 & 0.003 &  0.013  &  0.003    \\
      Reinteractions of secondary particles & 0.014 & 0.070  &  0.017 &  0.067  &  0.018    \\
      Beam position and divergence      &  0.056  &   0.021  &  0.058 &  0.035  &  0.002    \\
      Horn current                      &  0.004  &   0.004  &  0.001 &  0.001  &  0.005    \\
      Field in inner conductor          &  0.004  &   0.026  &  0.011 &  0.016  &  0.007    \\
      Amount of material                &  0.012  &   0.022  &  0.007 &  0.012  &  0.005    \\
      Horn misalignment                 &  0.002  &   0.006  &  0.007 &  0.012  &  0.005    \\
      Collimator misalignment           &  0.003  &   0.020  &  0.008 &  0.013  &  0.005    \\ \hline
      Total                             &  0.068  &   0.091  &  0.074 &  0.103  &  0.042    \\ \hline
    \end{tabular}
  \end{center}
  \label{tab:normerrors}
\end{table}

In order to check the effect of the functional form used in the fits to
the SPY/NA20 points, different order polynomials were tried.  The resulting
envelope of 3000 simulated experiments was essentially the same as the
original one and therefore no additional uncertainty was assigned from
this source.

\subsection{Systematic uncertainties from the yields of particles in
  interactions other than p-Be}
\label{sec:syst_reint}
As described in Section~\ref{sec:mc}, the interactions of protons downstream
of the target, the reinteractions of particles downstream of the target
and the reinteractions of particles in the target were treated by FLUKA 2000,
in the first two cases by correcting the GFLUKA estimates by
the ratio between FLUKA 2000 and GFLUKA and in the third case
by treating them directly with FLUKA 2000.  The yields of mesons
from these three sources could not be corrected by the reweighting factors
obtained from the SPY and NA20 measurements since these experiments did not
measure proton interactions in materials other than beryllium nor
interactions of particles other than protons.  However the reweighting
factors discussed in Section~\ref{sec:reweight} were used to estimate
the uncertainty on the neutrino fluxes from these three sources as
explained below.

For each produced meson type two quantities were defined on the basis of
the reweighting factors shown in Fig.~\ref{fig:fitExample}.  $D_{\rm max}$
was the maximum deviation from unity of the reweighting factor between
20 and 100 GeV/$c$ and $D_{\rm ave}$ was the average deviation from unity
in the same momentum range.  $D_{\rm max}$ was 10.0\%, 15.0\%, 17.0\%,
27.0\%, 26.6\% and 15.0\% for $\pip$, $\kp$, $\pim$, $\km$, $\klong$ and
``others'', respectively.  The corresponding values for $D_{\rm ave}$ were
5.8\%, 4.0\%, 15.3\%, 26.5\%, 22.6\% and 10.0\%.


\begin{itemize}
\item {\it Proton interactions downstream of the beryllium target.}
  The momentum spectrum of each meson type resulting from these interactions
  was modified by $D_{\rm ave}$ for this meson type. The effect of this
  modification on the integral flux of each neutrino flavour was calculated.
  The effects from all meson types were added in quadrature and were
  included as a normalization error on each neutrino flavour and on the
  $\emurat$ ratio (line 2 of Table~\ref{tab:normerrors}).  Since these
  neutrinos affected the overall neutrino spectra similarly at all energies,
  the error was included wholly as a normalization error. \vspace{0.5cm}

\item {\it Reinteractions in the beryllium target and downstream of the
  target.} These two sources were treated separately but their errors were
  added linearly since they are correlated.  The contributions of these
  sources to the overall neutrino fluxes are very energy dependent and
  therefore the uncertainties were split into a normalization and an
  energy-dependent part.  The normalization uncertainty was calculated
  in the same way as for proton interactions downstream of the target
  and is shown in line 3 of Table~\ref{tab:normerrors}.  For the
  energy-dependent part, $D_{\rm ave}$ was subtracted in quadrature
  from $D_{\rm max}$, yielding $D_{\rm edep}$, resulting in values
  of $D_{\rm edep}$ of 8.1\%, 14.5\%, 7.4\%, 5.2\%, 14.0\% and 11.2\%
  for $\pip$, $\kp$, $\pim$, $\km$, $\klong$ and ``others'', respectively.
  The momentum spectrum of each meson type resulting from these
  reinteractions was then modified by $D_{\rm edep}$ and the effect
  of this modification was propagated to the energy spectrum of each
  neutrino flavour and to the $\emurat$ ratio.  The effect of all meson
  types on each neutrino flavour and on the ratio were combined in
  quadrature and were included as an energy-dependent uncertainty.
\end{itemize}

\subsection{Systematic uncertainties: other sources}
\label{sec:syst_others}
\begin{itemize}
\item {\it Position and angular divergence of the proton beam.}
  The $\pm 1\sigma$ uncertainty in the position of the beam relative
  to the target (measured by the beam scanner described in
  Section~\ref{sec:monitoring}) was
  $\pm 0.25$~mm.  The effect of this uncertainty on the normalization
  uncertainties (listed in line 4 of Table~\ref{tab:normerrors})
  amounted to 5.6\% on $\numu$ and 5.8\% on $\nue$.  It also produced
  an energy-dependent error of up to 2.3\% on the $\emurat$ ratio.
  The effect of the uncertainty on the angular divergence of the beam
  on neutrino fluxes was found to be negligible.
\end{itemize}

All subsequent uncertainties were included as normalization uncertainties.

\begin{itemize}
\item {\it Magnetic field in the horn and the reflector.}  The effect
  of the uncertainty in the magnetic field of the focusing elements on
  neutrino fluxes was studied by varying the nominal current value used
  in the simulation by $\pm 2$\%, the tolerance limit of the on-line control
  system, and noting the corresponding changes in the neutrino fluxes
  at NOMAD.  These changes were 0.4\% for $\numu$, 0.1\% for $\nue$
  and 0.5\% for $\emurat$.  We also studied the effect of the
  uncertainties in the knowledge of the magnetic field inside the
  inner conductor of the horn; the numbers obtained were 0.4\% for
  $\numu$, 1.1\% for $\nue$ and 0.7\% for $\emurat$. \vspace{0.5cm}

\item {\it Inaccuracies in the simulation of the beam line elements.}
  The size of these inaccuracies was estimated by studying the
  differences between the measured and predicted spectra of $\anumu~\cc$
  and $\anue~\cc$ events, which are the most sensitive to the secondary
  interactions in the beam elements (see Section~\ref{sec:mat}).  We
  found that the amount of material possibly missing in the
  simulation of the beam line does not exceed the equivalent of
  a slab of aluminium 1~cm thick, located downstream of the focusing elements.
  Increasing the amount of material in the beam line by this amount
  in the Monte Carlo simulation changed
  the expected $\numu$ flux by 1.2\%, the $\nue$ flux by 0.7\%,
  and the $\emurat$ ratio by 0.5\%. \vspace{0.5cm}

\item {\it Misalignment of the beam line elements.}  We have studied
  the effects of possible misalignments of the horn and of the aluminium
  collimator.  The upper limits on the misalignment of the horn, 1~mm
  in the horizontal and 1~mm in the vertical direction, were obtained
  by comparing the measured spatial distribution of $\numu~\cc$
  events with the results of several Monte Carlo simulations for
  various horn displacements with respect to its ideal position.
  The effect of this uncertainty on neutrino fluxes was 0.2\% for
  $\numu$, 0.7\% for $\nue$ and 0.5\% for $\emurat$.  The uncertainty
  in the collimator position (3~mm in both horizontal and vertical
  directions) gave rise to a 0.3\% uncertainty in the $\numu$ flux,
  0.8\% in the $\nue$ flux and 0.5\% in the $\emurat$ ratio. 
\end{itemize}

\subsection{Summary of systematic uncertainties}\label{sec:syst_summary}
\begin{figure}
\begin{center}
  \includegraphics*[scale=0.69]{\epsdir/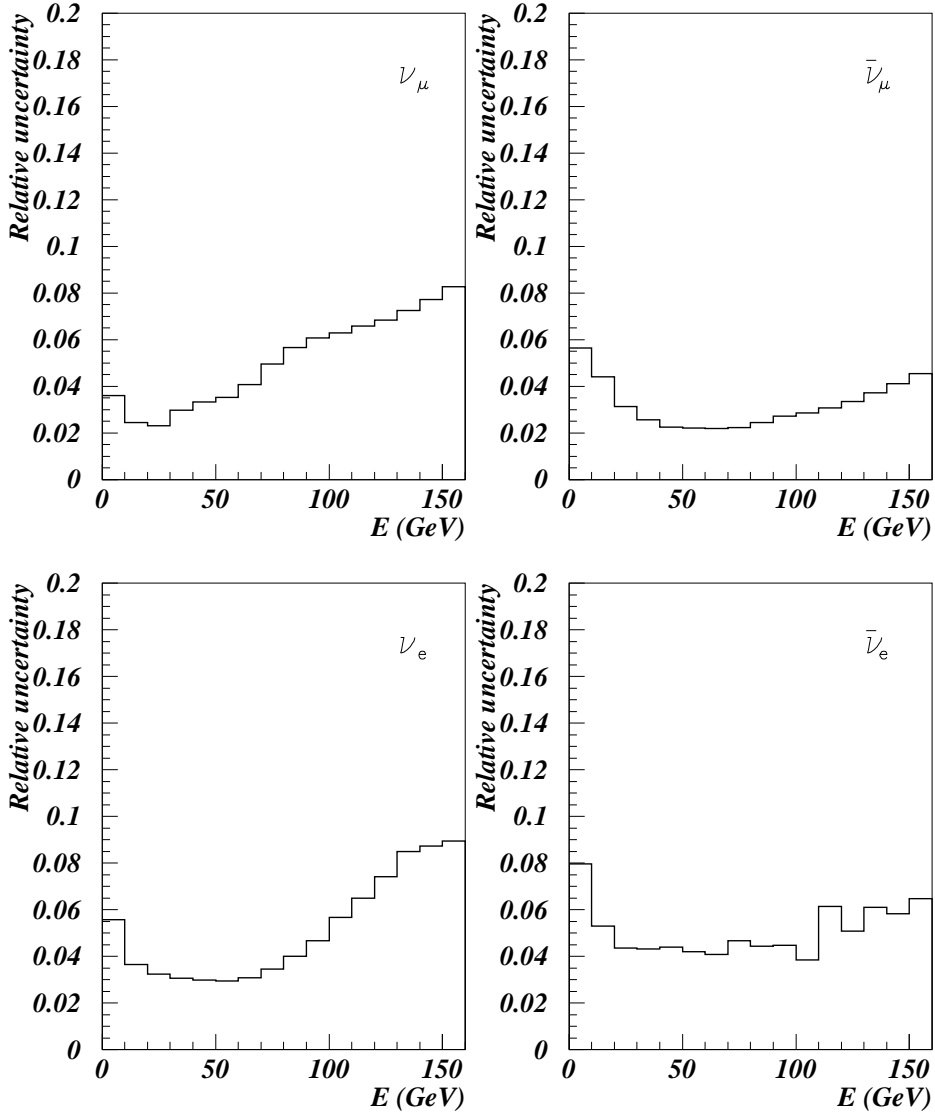}
\end{center}
\caption{Total energy-dependent uncertainties on the yields of each
  of the four neutrino species.  The energy-independent uncertainties
  are listed in Table~\ref{tab:normerrors}.}
\label{fig:err_nus}
\end{figure}

\begin{figure}
\vspace*{0.3cm}
\begin{center}
  \includegraphics*[scale=0.69]{\epsdir/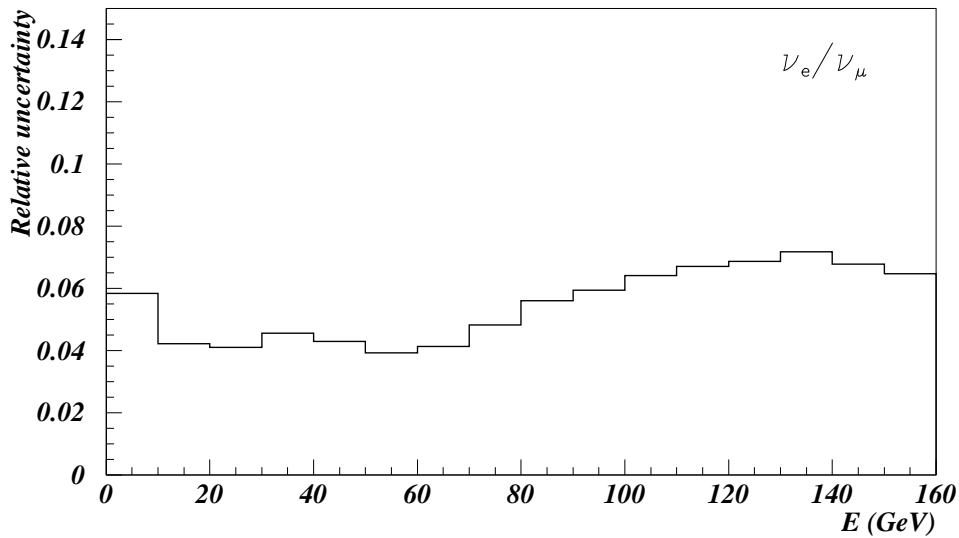}
\end{center}
\vspace*{-0.2cm}
\caption{Total energy-dependent uncertainty on the $\emurat$ ratio.
  The energy-inde\-pendent uncertainty is given in Table~\ref{tab:normerrors}.}
\label{fig:err_ratio}
\end{figure}

The overall energy-dependent uncertainties are shown in Fig.~\ref{fig:err_nus}
for the four neutrino species and in Fig.~\ref{fig:err_ratio} for
the $\emurat$ ratio.  The normalization systematic uncertainties
are summarized in Table~\ref{tab:normerrors}.

It should be noted that the normalization uncertainties of the $\numu$
and $\nue$ components of the beam could be reduced significantly through
a better knowledge of the beam position or through the use of a wider
target that would minimize the number of protons missing it.

\section{NOMAD apparatus and running conditions}
\label{sec:NOMAD}

\subsection{Detector}
\label{sec:detector}
 The NOMAD detector~\cite{NOMADNIM} consisted of a number of subdetectors
 most of which were located inside a large dipole magnet
 delivering a field of 0.4~T. The direction of the field was horizontal
 and perpendicular to the neutrino beam.

 An array of scintillator counters, $V$, covered the front face of the
 magnet and was used to veto interactions caused by muons accompanying
 the neutrino beam.  An active target consisting of 132 planes of drift
 chambers~\cite{DC} of 3$\times$3~m$^2$ occupied the upstream part of
 the magnet.  The fiducial mass of 2.7~tons was provided by
 the walls of the drift chambers. The average density of the active
 target of 0.1 g/cm$^3$ was low enough to allow accurate measurements
 of the individual particles produced in the neutrino interactions and
 to minimize their reinteractions. The momentum resolution for an average
 track length of 1.5~m was 4\% at 1~GeV/$c$ rising to 15\% at 50~GeV/$c$.

 The chambers were followed by 9 transition radiation (TRD) modules~\cite{TRD}
 for electron-pion discrimination. Each module consisted of a radiator
 of poly\-pro\-pylene foils followed by a detection plane of straw tubes.
 The TRD yielded a pion rejection factor of 1000 for an electron
 efficiency of 90\% in the momentum range 1 to 50~GeV/$c$. Two
 scintillation counter trigger planes~\cite{TRIG}, $T_1$ and $T_2$, bracketed
 the TRD.

 A lead glass array~\cite{ECAL} was located at the end of the magnet.
 It measured the energies and directions of photons and electrons with
 a resolution of
\begin{equation}
 \Delta E/E = (1.04 \pm 0.01)\% + (3.22 \pm 0.07)\% / \sqrt{E ({\rm GeV})}.
\end{equation}
 It was preceded by a preshower consisting of a 1.6 X$_\circ$ lead plate
 followed by two planes of proportional tubes. It was used for better
 photon localization and for further electron-pion discrimination.

 An iron-scintillator hadronic calorimeter was located outside the magnet
 and was followed by two stations of drift chambers for muon identification.
 The first station was located after 113~cm of iron and the second after
 an additional 80~cm of iron. This allowed identification of muons
 with momentum larger than 2.5~GeV/$c$.

 With the detectors described above, NOMAD had excellent electron and
 muon identification and therefore could reconstruct and identify $\numu~\cc$,
 $\anumu~\cc$, $\nue~\cc$ and $\anue~\cc$ interactions.

\subsection{Running conditions}
\label{sec:dtps}
 NOMAD collected data from 1995 to 1998. The main trigger, $\bar{V} \times
 T_1 \times T_2$, consisted of a coincidence between signals from the two
 trigger planes in the absence of a signal in the entrance
 veto detector. It was designed to record interactions of neutral particles
 in the target. Most of the running, a total exposure of $5.1 \times 10^{19}$
 protons on target (p.o.t.), was in neutrino mode and yielded about
 $1.3 \times 10^6$ $\numu~\cc$ interactions. Some data, amounting
 to $0.44 \times 10^{19}$ p.o.t., were also collected in antineutrino
 mode (reverse polarity in the horn and reflector) and some,
 $0.04 \times 10^{19}$ p.o.t., in zero-focusing mode (with the horn
 and reflector switched off); these data were used mostly to check
 the beam line simulation.  In particular, the ability of our simulation
 programs to reproduce the energy spectra of neutrinos of different
 species at all three settings of the horn demonstrated that the magnetic
 field of the horn and the material in the beam line were well simulated.

\section{Comparison with data}
\label{sec:compData}
The results of our simulations of neutrino fluxes were compared with
the data collected in NOMAD.  For this purpose we have generated large
samples of $\numu$, $\anumu$, $\nue$ and $\anue$ interactions in the
NOMAD detector according to the energy spectra and radial distributions
calculated for each neutrino species.  Our event generator included
deep-inelastic, quasi-elastic and resonance events, and was complemented
by a full simulation of the detector response based on GEANT~3.21~\cite{GEANT}.
A detailed description of the NOMAD simulation is given in
Ref.~\cite{NUTAU_LATEST}.  The energy of the hadronic system produced
in a neutrino interaction was reduced~\cite{NUE_LATEST} by 8.3\% in the
Monte Carlo in order to account for losses in the data not very well
described by our detector simulation.

Two sets of selection criteria were applied to both the data
and the Monte Carlo events.  The first set of cuts selected events
with a prompt isolated muon in the final state; depending on the sign
of the muon, these events were classified as $\numu~\cc$ or $\anumu~\cc$
interactions.  The second set of cuts selected events with a prompt
isolated electron or positron (and no muon); these events were classified
as $\nue~\cc$ and $\anue~\cc$ interactions, respectively.  The non-prompt
background contaminations, mainly from pion and kaon decays in the muon
sample and from photon conversions in the electron sample, were evaluated
and taken into account.  Including the small contribution from the wrong
charge assignment to the lepton, the fractions of background amounted
to 0.1\% for $\numu~\cc$, 15.2\% for
$\anumu~\cc$, 2.3\% for $\nue~\cc$ and 32.2\% for $\anue~\cc$ samples
for the neutrino mode.  A detailed description of the selection
of events used in the comparison can be found in Ref.~\cite{NUE_LATEST}.
The summary of all available data samples is given in
Table~\ref{tab:events_stat}.

\begin{table}[htb]
 \caption{Number of observed $\numu~\cc$, $\anumu~\cc$, $\nue~\cc$
   and $\anue~\cc$ events after selection cuts, in neutrino, antineutrino,
   and zero-focusing modes.}
  \begin{center}
    \begin{tabular}{crrrr} \hline
                    & $\numu~\cc$ & $\anumu~\cc$ & $\nue~\cc$ & $\anue~\cc$ \\ \hline
 Neutrino mode      &    830,535  &     27,646   &     --     &     1,446   \\
 Antineutrino mode  &      8,176  &     26,996   &     245    &       267   \\
 Zero-focusing mode &      1,661  &        534   &      35    &        10   \\ \hline
    \end{tabular}
  \end{center}
  \label{tab:events_stat}
\end{table}

\begin{figure}
\begin{center}
  \includegraphics*[scale=0.69]{\epsdir/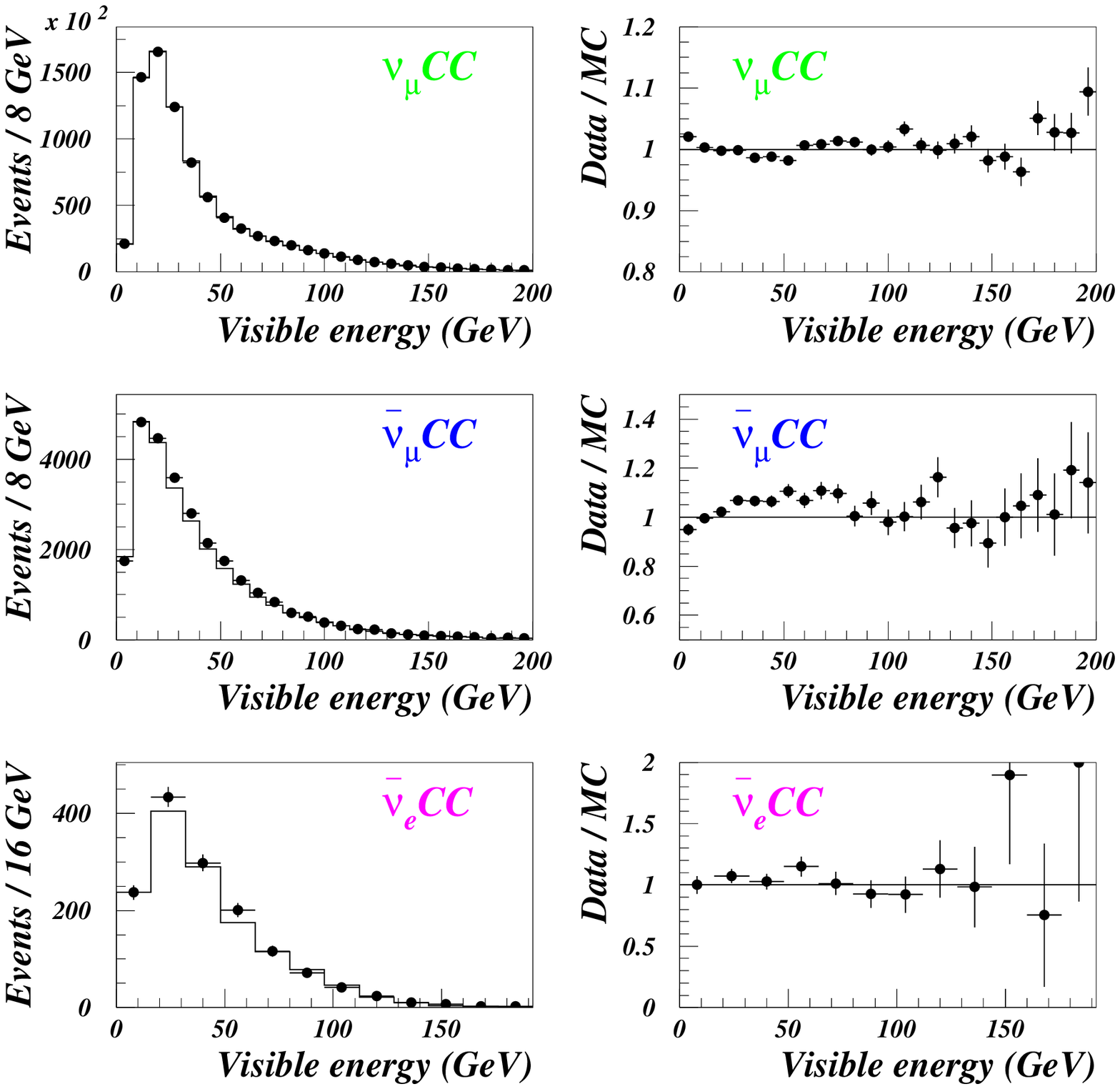}
\end{center}
\vspace*{-0.4cm}
\caption{Left: neutrino energy spectra for the data (points with error
  bars) and the Monte Carlo (histogram), for $\numu~\cc$ (top),
  $\anumu~\cc$ (middle) and $\anue~\cc$ (bottom) interactions
  in {\it neutrino mode}.  Right: ratios of the measured to
  the predicted distributions, for the same three neutrino species.
  The errors shown are statistical only.}
\label{fig:foc+}
\end{figure}

\begin{figure}
\vspace*{0.7cm}
\begin{center}
  \includegraphics*[scale=0.69]{\epsdir/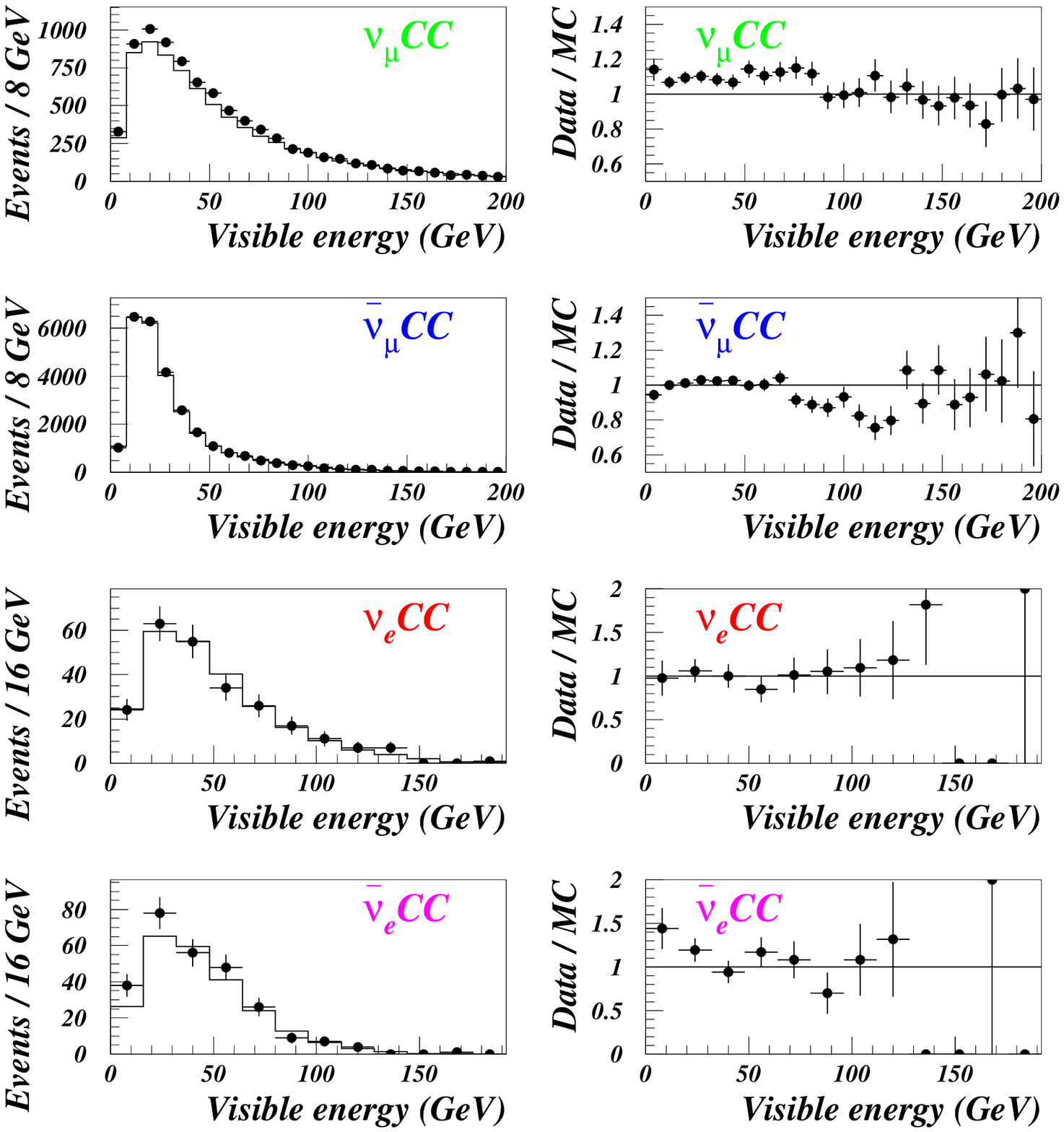}
\end{center}
\caption{Left: neutrino energy spectra for the data (points with error
  bars) and the Monte Carlo (histogram), for (from top to bottom)
  $\numu~\cc$, $\anumu~\cc$, $\nue~\cc$ and $\anue~\cc$
  interactions in {\it antineutrino mode}.  Right: ratios
  of the measured to the predicted distributions, for the same four
  neutrino species.  The errors shown are statistical only.}
\label{fig:foc-}
\end{figure}

In Fig.~\ref{fig:foc+} we show the comparison between the measured
and the predicted neutrino energy spectra for $\numu~\cc$, $\anumu~\cc$
and $\anue~\cc$ events in neutrino mode.  The corresponding
comparison for $\nue~\cc$ interactions cannot be shown here as it has been
the subject of a search for \nmne oscillations using a ``blind''
analysis\footnote{The $\anue~\cc$ spectrum can be shown since, even if
there were oscillations within the allowed parameter space, their effect
would not be very visible in this spectrum given that the intrinsic
$\anumu/\anue$ ratio of the beam is four times smaller than the intrinsic
$\numu/\nue$ ratio (Table~\ref{tab:fluka_comp}) and given the limited
antineutrino statistics.};
it is discussed in a separate paper~\cite{NUE_LATEST}.
The neutrino energy was approximated by the ``visible energy'', defined
as the sum of the energies of the charged lepton and of the hadrons
observed in the final state.  Since the main purpose of this detailed
prediction of neutrino fluxes is the study of \nmne oscillations using
the $\emurat$ ratio, it was sufficient to normalize the Monte Carlo
distribution of $\numu~\cc$ events to the number of $\numu~\cc$ events
in the data.  Hence, only the shape of the $\numu~\cc$ distributions
can be compared; nonetheless it is noteworthy that the shape of the
$\numu~\cc$ energy spectrum is predicted to better than 2\% up to
150~GeV. For the normalization of $\anumu~\cc$ and $\anue~\cc$ simulated
events we use the relative $\anumu$/$\numu$ and $\anue$/$\numu$
abundances predicted by our simulation.  Therefore both the number
of events and the shape of the spectra can be compared.  The comparison
shows that the results of our simulations are in very good agreement
with the data.  The only statistically significant
difference between the data and the Monte Carlo predictions is
a difference of up to about 8\% in the expected number of $\anumu~\cc$
events; this difference is smaller than the estimated uncertainty
of our $\anumu$ flux predictions.  Both the shape and the total number
of $\anue~\cc$ events are well reproduced.  This confirms the validity
of our description of the yields of $\klong$ (the principal source of
$\anue$) and of our estimates of the
background contamination from processes other than $\anue~\cc$ interactions.

\begin{figure}
\vspace*{-0.3cm}
\begin{center}
  \includegraphics*[scale=0.69]{\epsdir/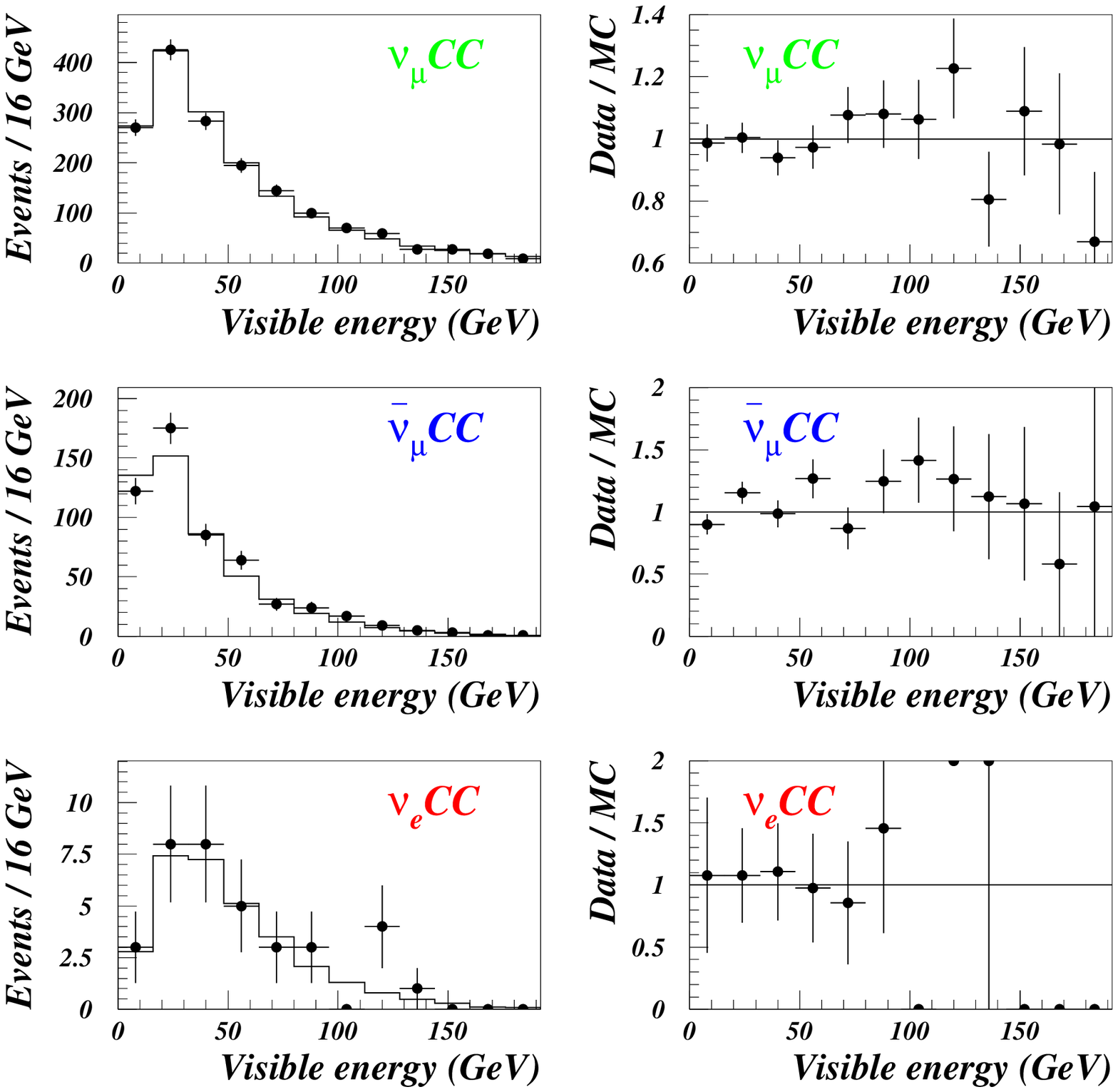}
\end{center}
\vspace*{-0.4cm}
\caption{Left: neutrino energy spectra for the data (points with error
  bars) and the Monte Carlo (histogram), for $\numu~\cc$ (top),
  $\anumu~\cc$ (middle) and $\nue~\cc$ (bottom) interactions
  in {\it zero-focusing mode}.  Right: ratios of the measured to
  the predicted distributions, for the same three neutrino species.
  The errors shown are statistical only.}
\label{fig:foc0}
\end{figure}

The comparison between the measured and the predicted neutrino energy
spectra for $\numu~\cc$, $\anumu~\cc$, $\nue~\cc$ and $\anue~\cc$
events in antineutrino mode is shown in Fig.~\ref{fig:foc-}.
Similarly, the comparison between the data and the Monte Carlo for
zero-focusing mode is shown in Fig.~\ref{fig:foc0}.
The Monte Carlo distributions of the most abundant neutrino flavour
in each data-taking mode ($\anumu$ in antineutrino and $\numu$ in
zero-focusing) are again normalized to the total number of corresponding
events in the data; the predicted distributions of all other species
are normalized using their relative abundances predicted by our simulation.
The good agreement between the data and the Monte Carlo is an
important confirmation of the validity of the beam line simulation.

\begin{figure}
\begin{center}
  \includegraphics*[scale=0.69]{\epsdir/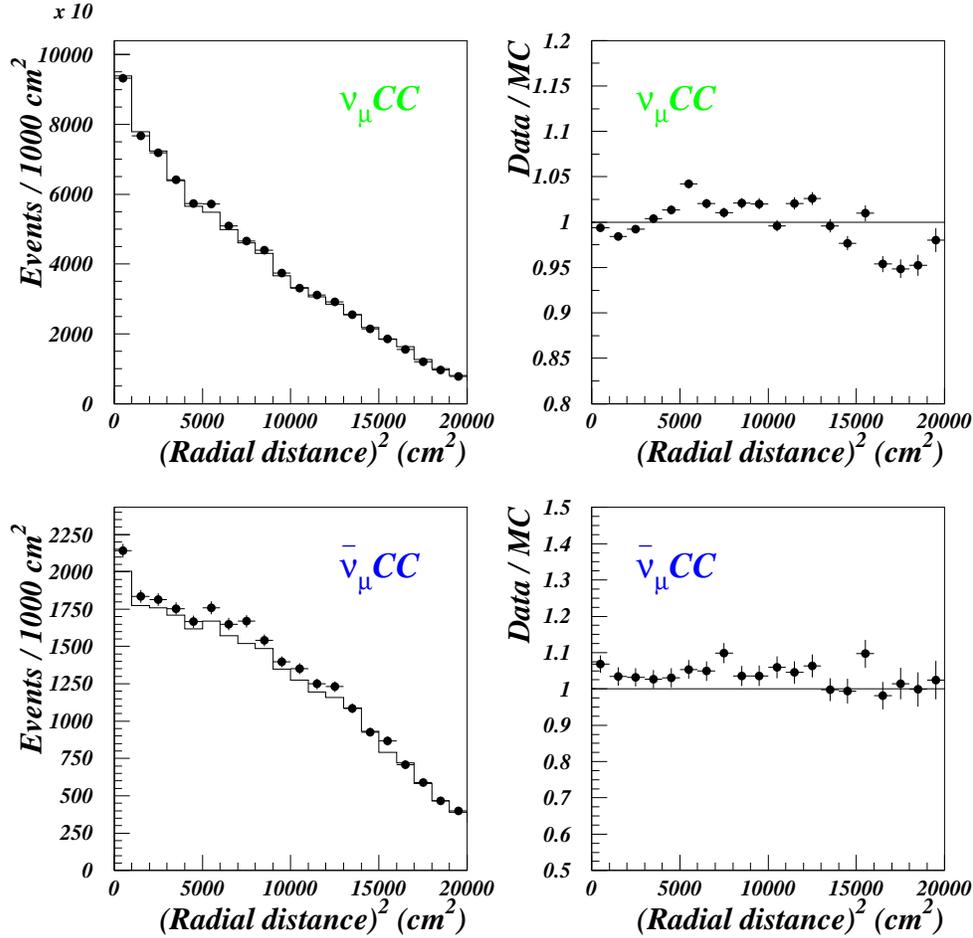}
\end{center}
\vspace*{-0.4cm}
\caption{Left: distributions of the square of the radial position
  of neutrino interaction vertex for the data (points with error
  bars) and the Monte Carlo (histogram), for $\numu~\cc$ (top)
  and $\anumu~\cc$ (bottom) interactions in neutrino mode.
  Right: ratios of the measured to the predicted distributions.
  The errors shown are statistical only.}
\label{fig:rad_foc+}
\end{figure}

Finally, Fig.~\ref{fig:rad_foc+} shows the comparison between the
measured and the simulated radial distributions of the neutrino
interaction vertex for $\numu~\cc$ and $\anumu~\cc$ events in neutrino
mode. The radial position of each interaction was calculated
with respect to the nominal beam axis. The predictions agree with the data
to better than
5\%.  Both the energy and the radial dependence of the $\emurat$
ratio are used in the search for \nmne oscillations, substantially
increasing the sensitivity of the search.

\section{\boldmath Prediction of the $\emurat$ ratio}
\label{sec:ratio}
\begin{figure}[htb]
\vspace*{-0.3cm}
\begin{center}
  \includegraphics*[scale=0.69]{\epsdir/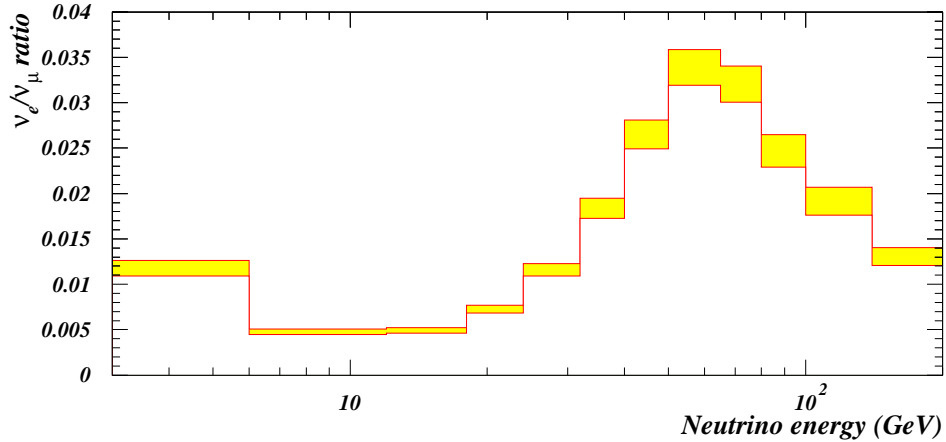}
\end{center}
\vspace*{-0.3cm}
\caption{Ratio $\emurat$ as a function of neutrino energy at NOMAD, within
  the transverse fiducial area of 260$\times$260~cm$^2$.  The upper and
  lower boundaries of the filled band correspond to the predictions with
  $\pm 1\sigma$ uncertainty, where $\sigma$ includes both the normalization
  and energy-dependent uncertainties added in quadrature.}
\label{fig:nue_numu_rat}
\end{figure}

The most probing test of this beam simulation is the prediction of the
$\emurat$ ratio, which is shown in Fig.~\ref{fig:nue_numu_rat}.
The corresponding NOMAD data will be shown in a forthcoming paper,
Ref.~\cite{NUE_LATEST}, on the search for \nmne oscillations.

\section{Conclusions}
\label{sec:conclusions}
A detailed simulation of the WANF neutrino beam has been developed by
the NOMAD collaboration in order to
predict the flavour content of this beam.  The simulation was based on
particle yields calculated using the FLUKA package.  These
yields were adapted to agree with the data of the NA20 and SPY particle
production experiments.  The fluxes of the four neutrino flavours at NOMAD
were predicted with an overall uncertainty of about 8\% for $\numu$
and $\nue$, 10\% for $\anumu$, and 12\% for $\anue$ (energy-dependent
and normalization errors combined).

The main purpose of this detailed simulation was the prediction of the
$\emurat$ ratio for the search for \nmne oscillations.  The energy-dependent
uncertainty achieved on this prediction ranges from 4 to 7\%
whereas its normalization uncertainty is 4.2\%.

\begin{ack}
The following funding agencies have contributed to this experiment:
Australian Research Council (ARC) and Department of Industry, Science, and
Resources (DISR), Australia;
Institut National de Physique Nucl\'eaire et Phy\-sique des Particules (IN2P3),
Commissariat \`a l'Energie Atomique (CEA), Mi\-nist\`ere de l'Education
Nationale, de l'Enseignement Sup\'erieur et de la Re\-cherche, France;
Bundesministerium f\"ur Bildung und Forschung (BMBF),
Germany;
Istituto Nazionale di Fisica Nucleare (INFN), Italy;
Institute for Nuclear Research of the Russian Academy of Sciences, Russia;
Fonds National Suisse de la Recherche Scientifique, Switzerland;
Department of Energy, National Science Foundation, 
the Sloan and the Cottrell Foundations, USA.

We thank the management and staff of CERN and of all
participating institutes for their vigorous support of the experiment.
Particular thanks are due to the CERN accelerator and beam-line staff
for the magnificent performance of the neutrino beam.  We are especially
grateful to V.~Falaleev, J.-M.~Maugain, S.~Rangod and P.R.~Sala for
their invaluable contribution to the design and operation of the WANF
and for their help in the simulation of the WANF beam line.
We also thank our secretarial staff, J.~Barney, K.~Cross,
J.~Hebb, M.-A.~Huber, N.~Marzo, J.~Morton, R.~Phillips and M.~Richtering,
and the following people who have worked with the
collaboration on the preparation and the data
collection stages of NOMAD:
M.~Anfreville, M.~Authier, G.~Barichello, A.~Beer, V.~Bonaiti, A.~Castera,
O.~Clou\'e, C.~D\'etraz, L.~Dumps, C.~Engster,
G.~Gallay, W.~Huta, E.~Lessmann,
J.~Mulon, J.P.~Pass\'e\-ri\-eux, P.~Petit\-pas, J.~Poin\-signon,
C.~Sob\-czyn\-ski, S.~Sou\-li\'e, L.~Vi\-sen\-tin, P.~Wicht.
Finally we acknowledge the fruitful collaboration with our
colleagues from CHORUS during the setting-up, monitoring and understanding
of the beam line.
\end{ack}

\newcommand\bff{}

\end{document}